\documentclass[lettersize,journal]{IEEEtran}
\usepackage{amsmath,amsfonts}
\usepackage{algorithmic}
\usepackage{algorithm}
\usepackage{array}
\usepackage[caption=false,font=normalsize,labelfont=sf,textfont=sf]{subfig}
\usepackage{caption}
\usepackage{textcomp}
\usepackage{stfloats}
\usepackage{url}
\usepackage{verbatim}
\usepackage{graphicx}
\usepackage{subcaption}
\usepackage{cleveref}
\usepackage{cite}
\usepackage{mathtools}
\usepackage{pifont}
\usepackage{amsmath,amssymb,amsfonts}
\usepackage{amsthm,amsmath,amssymb}
\usepackage{mathrsfs}
\usepackage{enumerate}
\usepackage{subcaption}
\usepackage{times}
\usepackage{float}
\usepackage{color}
\usepackage{cuted}
\newtheorem{theorem}{Theorem}
\newtheorem{corollary}{Corollary}[theorem]
\usepackage{xpatch}  
\makeatletter  
\xpatchcmd{\@thm}{\thm@headpunct{.}}{\thm@headpunct{}}{}{}
\makeatother  
\begin{document}

\title{Beamforming-based Achievable Rate Maximization in ISAC System for Multi-UAV Networking }
%Temporal-Assisted Beamforming and Prediction in Sensing-Enabled UAV Communications

\author{\IEEEauthorblockN{
            Shengcai Zhou, \emph{Student~Member, IEEE},
			Luping Xiang, \emph{Senier Member, IEEE},
            Kun Yang, \emph{Fellow, IEEE},
            Kai Kit Wong, \emph{Fellow, IEEE},
            Dapeng Oliver Wu, \emph{Fellow, IEEE},
            and Chan-Byoung Chae, \emph{Fellow, IEEE}
             }
			\vspace{-0.5 cm}\\

%\thanks{This work was supported in part by the Natural Science Foundation of China under Grant 62301122 and Grant 62071101; in part by the Fundamental Research Funds for the Central Universities under Grant ZYGX2019J001; in part by the Sichuan Science and Technology Program under Grant 2023NSFSC1375. \textit{(Corresponding author: Luping Xiang.)}}
        \thanks{S. Zhou is with the Yangtze Delta Region Institute (Quzhou), University of Electronic Science and Technology of China, Quzhou 324003, China, and also the School of Information and Communication Engineering, University of Electronic Science and Technology of China, Chengdu 611731, China, email: 202222011112@std.uestc.edu.cn.}
        \thanks{L. Xiang and K. Yang are with the State Key Laboratory of Novel Software Technology, Nanjing University, Nanjing 210008, China, and School of Intelligent Software and Engineering, Nanjing University (Suzhou Campus), Suzhou 215163, China, email: luping.xiang@nju.edu.cn, kunyang@nju.edu.cn.}
        \thanks{K.-K. Wong is with the Department of Electronic and Electrical Engineering, University College London, WC1E 7JE London, U.K., and also with the Yonsei Frontier Laboratory, Yonsei University, Seoul 03722, South Korea (e-mail: kai-kit.wong@ucl.ac.uk).}
        \thanks{D. O. Wu is with the Department of Computer Science, City University of Hong Kong, Hong Kong, e-mail: dpwu@ieee.org.}
        \thanks{C.-B. Chae is with the School of Integrated Technology, Yonsei University, Seoul 03722, South Korea (e-mail: cbchae@yonsei.ac.kr).}
	}

\maketitle

% The paper headers
%\markboth{Journal of \LaTeX\ Class Files,~Vol.~14, No.~8, August~2021}%
%{Shell \MakeLowercase{\textit{et al.}}: A Sample Article Using IEEEtran.cls for IEEE Journals}

%\IEEEpubid{0000--0000/00\$00.00~\copyright~2021 IEEE}
% Remember, if you use this you must call \IEEEpubidadjcol in the second
% column for its text to clear the IEEEpubid mark.

\maketitle

\begin{abstract}
Integrated Sensing and Communication (ISAC) combined with Unmanned Aerial Vehicle (UAV) networking offers a promising solution for resilient wireless connectivity in post-disaster scenarios. In such settings, UAVs must rapidly explore unknown environments while providing continuous sensing-aided communication services. This paper proposes a novel ISAC framework based on a temporal-assisted frame structure that integrates search, tracking, and communication functions for multi-UAV networks. We formulate a joint optimization problem that maximizes the total achievable rate per slot through the coordinated design of UAV beamforming, load management, and UAV direction planning, while accounting for realistic sensing uncertainties. The NP-hard problem is decomposed into three sub-problems and solved through a combination of an enhanced distributed Successive Convex Approximation (SCA)-Iterative Rank Minimization (IRM) algorithm, a coalition game-based load allocation strategy, and a Fermat point-based UAV path planning method. Simulation results demonstrate that the proposed scheme significantly improves communication capacity, fairness, and sensing accuracy compared to conventional UAV-assisted ISAC designs. This work provides practical design insights for robust and efficient ISAC-based UAV networks.
\end{abstract}


\section{Introduction}
%\IEEEPARstart{T}{his}

\subsection{Background}
\IEEEPARstart{T}{o} achieve the vision of the Internet of Everything (IoE), Integrated Sensing and Communication (ISAC) technology has been identified as a key innovation by the International Telecommunication Union Radiocommunication Sector (ITU-R) for 6G wireless communication \cite{10601686}. Recently, various ISAC-based schemes have been continuously proposed, including but not limited to scenarios such as the Internet of Vehicles (IoV), telemedicine, and smart factories \cite{9737357,10158711,peng2025simac}.

Unmanned aerial vehicles (UAVs) are poised to play a crucial role in future wireless communication networks \cite{9729746,9282206,10529184,10791445}, primarily because they can compensate for the lack of ground-based ISAC systems, particularly in post-disaster scenarios \cite{7995044,8660516,8918497,10806828}. Ground obstacles often obstruct the line-of-sight (LoS) links between ground stations and remote targets, whereas UAVs, operating at higher altitudes, can maintain strong air-to-ground LoS connections \cite{9916163,10098686,10680056}.

The move toward high-frequency operation implies that communication will increasingly share the millimeter-wave (mmWave) band with radar \cite{Fan2020RadarcommunicationSS}. Given the need for high-speed communication and high-precision positioning, adapting millimeter-wave and massive multiple-input multiple-output (mMIMO) technologies, commonly used in ground stations, for UAV platforms is a promising direction. MmWave offers high bandwidth and fine range resolution, while mMIMO array gain mitigates path loss and enhances angular resolution \cite{8246850}. Moreover, UAVs can be flexibly deployed, either statically within an area of interest to identify potential targets, or dynamically according to real-time operational needs such as path planning, thereby optimizing ISAC performance \cite{10529184}. Additionally, ISAC enables the use of a shared waveform on common hardware for both sensing and communication, reducing the UAV's payload \cite{10529184}, which presents exciting prospects for UAV-enabled ISAC.

Furthermore, the sensing coverage, communication rate, and application scenarios of an individual UAV are inherently limited. Employing multiple UAVs for ISAC services can effectively extend coverage and enhance integration gains~\cite{9293257}. For instance, through information sharing and data fusion, UAVs can eliminate redundant target detections and improve tracking accuracy \cite{10098686}. However, despite considerable efforts to enhance multi-UAV network performance in terms of both communication~\cite{8247211,8770103} and sensing~\cite{7452646,7472256}, optimizing ISAC-based UAV networks remains a complex challenge due to the involvement of multiple targets.

\subsection{Related Work}
We first review the recent achievements of ISAC in vehicle-to-everything (V2X) \cite{7888145,9171304,9246715,10061429,9947033,9557830,9945983}. In recent years, there has been a lot of research on sensing-assisted communication. One typical approach involves equipping roadside unit (RSU) with dedicated radar sensors to enhance vehicle-to-infrastructure (V2I) communication beamforming \cite{7888145}. While this can improve communication performance and reduce overhead, it also requires additional hardware resources. Another scheme \cite{8851151} improves the traditional mmWave channel estimation and tracking framework by reducing the number of uplink feedback symbols. In fact, in mobility scenarios, relying solely on beam tracking is insufficient; UAVs must possess predictive capabilities to ensure beam alignment and mitigate potential communication disruptions. To optimize hardware usage, \cite{9171304} proposed a sensing-assisted predictive  beamforming method, which utilizes integrated mmWave echoes for beam alignment, realizing simultaneous communication and sensing services. Compared with the traditional beam alignment scheme \cite{8809900} and\cite{7999215}, \cite{9171304} eliminates the need for dedicated downlink pilots and uplink feedback, avoids quantization errors, and adds an extra positioning dimension. In simple terms, by replacing the pilot signal, \cite{9171304} provides greater localization accuracy and higher achievable rates. 

To estimate motion parameters more accurately, a message passing algorithm based on a factor graph was proposed in \cite{9246715}, which offers lower complexity than the extended Kalman filtering (EKF) scheme. However, both \cite{9171304} and \cite{9246715} consider scenarios that may be less practical in real-world contexts, such as assuming vehicles travel on a straight road parallel to the antenna array, whereas in reality, road geometry is much more complex. This limitation motivated \cite{10061429}, where the researchers constructed a curvilinear coordinate system (CCS) to design beams for arbitrarily shaped roads. Furthermore, \cite{9171304} and \cite{9246715} also assumed the vehicle to be a point target, while \cite{9947033} considered extended targets and proposed new frame structures and dynamic beamforming schemes to achieve a trade-off between sensing and communication. In addition, ISAC-assisted orthogonal time-frequency space (OTFS) technology was proposed in \cite{9557830}. Leveraging the advantages of OTFS signals in high-speed scenarios, \cite{9557830} introduced a simplified frame structure. Meanwhile, ISAC performance indicators continue to improve, and \cite{9945983} has established a unified ISAC resource allocation framework that considers sensing service fairness and comprehensive optimization criteria in V2X.

Similarly, ISAC has had many recent achievements in the UAV field \cite{10529184,9916163,10098686,9858656,10659350,9293257}.
\cite{10529184} emphasized employing the Cram$\acute{\text{e}}$r-Rao bound (CRB) of sensing parameters as a theoretical estimation index and formulates a joint optimization problem to balance CRB and achievable rate. With the introduction of uniform linear array (ULA), beamforming is further considered in \cite{9916163}, where UAVs, acting as a dual-function access point in the air, simultaneously communicate with multiple users and perform radar sensing on areas of interest. Subject to \cite{9916163}, \cite{10098686} constructed a complex frame structure with co-design and time-division design to group targets based on their locations, thereby improving sensing efficiency and energy efficiency. Instead of deep coupling of communication and sensing functions in the above, \cite{9858656} proposes an integrated periodic sensing and communication (IPSAC) mechanism, where communication and sensing functions are performed separately within a frame, allowing for a flexible trade-off. \cite{10659350} takes into account the UAV sensing of the extended target and transplants the  scheme in \cite{9947033} into the air. However, its assumption that the UAV carries 80 antennas is not practical. Furthermore, in the wireless communication system of multi-UAV network, joint optimization is a common research direction. \cite{8247211} jointly optimized user scheduling and association, UAV trajectory and transmission power, in which user scheduling association intuitively reflected the collaboration between UAVs. \cite{9293257} considered ISAC signals and proposes a coalition game method to achieve user load balance of each UAV, but the networked UAVs remained static without considering path planning. 

In post-disaster scenarios, an omnidirectional beampattern must be frequently transmitted for user searches; however, existing frame structures primarily focus on user tracking without exploring the relationship between ISAC omnidirectional and directional waveforms. Furthermore, a frame structure for temporal-assisted communication leveraging sensing information has not yet been developed. Compared to two-dimensional scenarios, UAVs operate more frequently in complex three-dimensional spaces, necessitating the use of uniform planar arrays (UPA) for effective beamforming. Unlike networked road environments, sensing information is often inaccurate during the initial stage of post-disaster search. Unfortunately, the impact of sensing information on transmitted beams remains underexplored. Although there is substantial literature on ISAC-based UAV path planning \cite{9858656,9625578,9130055}, many proposed solutions are overly complex, task-specific, and lack portability.

\begin{table*}[t]
\centering
% \small
\caption{Contrasting Our Contributions To The State-Of-The-Art}
\setlength{\tabcolsep}{4mm}{
\begin{tabular}{l|c|c|c|c|c|c|c}
\hline %draw line
Contributions & \textbf{this work} & \cite{10529184} & \cite{9916163} &\cite{9293257} & \cite{8247211} &\cite{9171304,7888145} & \cite{10061429,9947033}
\\ \hline\hline
Omni-directional protocol & \ding{52} &  &  &  & & &  \\
\hline
Sensing-assisted & \ding{52} &  &   & & &\ding{51} &\ding{51}  \\
\hline
Beam coverage design & \ding{52} &    &\ding{51}  &  && &\ding{51} \\
\hline
Multi-UAV network & \ding{52} &    & & \ding{51} &\ding{51}& & \\
\hline
Load management & \ding{52} &    &  &\ding{51} &\ding{51}&  & \\
\hline
Direction planning & \ding{52} & \ding{51} & \ding{51}   & &\ding{51}& & \\
\hline
 % Integrated signal & \ding{52} & \ding{51} & \ding{51} &  & \ding{51}&&\ding{51}&\ding{51}  \\
% \hline
\end{tabular}
}
\vspace{-0.3 cm}
\label{contributions}
\end{table*}

\subsection{Our Contributions}
Inspired by the above, we develop a search and communication scheme of multi-UAV network in post-disaster emergency communication. The innovation of this work is shown in Table~\ref{contributions}, and the main contributions are summarized as follows:
\begin{itemize}
\item We propose an ISAC framework with integrated search-track-communication functionality based on an omnidirectional protocol for multi-UAV networks in post-disaster environments. To enable continuous sensing-aided communication, we formulate a joint UAV beamforming, load management, and UAV direction planning problem aimed at maximizing the total achievable rate per slot.

\item We develop a distributed optimization approach to efficiently solve the proposed NP-hard problem. Specifically, we propose an improved Successive Convex Approximation (SCA)-Iterative Rank Minimization (IRM) algorithm to enable scalable multi-UAV beamforming based on shared sensing information. For load optimization, we employ a coalition game framework to design a flexible and fair load allocation strategy. For UAV direction planning, we introduce a lightweight yet effective Fermat point-based method to enhance coverage and improve ISAC performance.

%The proposed NP-hard problem is decomposed into three sub-problems, and we develop a suboptimal yet efficient method to solve it. First, we address multi-UAV beamforming based on shared information and propose an improved Successive Convex Approximation (SCA)-Iterative Rank Minimization (IRM) algorithm to enable distributed computation and reduce the computational load on each UAV. For load optimization, we employ a coalition game approach to optimize the network load allocation strategy. For the direction planning sub-problem, we adopt a simple yet versatile Fermat point search algorithm.

\item We provide extensive simulation results that demonstrate the effectiveness of the proposed scheme across three key aspects: beam design, load management, and direction planning. Compared to conventional UAV-based ISAC methods, our framework achieves significant gains in total achievable rate, fairness, and sensing accuracy, while maintaining computational efficiency suitable for real-time UAV deployment.
\end{itemize}

The remainder of this paper is organized as follows: Section II outlines the system model; Section III details the formulation of the optimization problem; Section IV presents the proposed solutions; Section V showcases the numerical results; and Section VI concludes the discussion.

\textit{Notations:} In this paper, $(\cdot)^T$, $\left\| \cdot \right\|$, $(\cdot)^H$ and $\mathrm{diag}(\cdot)$  stands for transpose, modulus of a vector, Hermitian and diagonal matrix. \textbf{A} $\otimes$ \textbf{B} denotes the Kronecker product of two matrices \textbf{A} and \textbf{B}.

\section{System Model}
\begin{figure}[h] %htbp
\centerline{\includegraphics[width=0.45\textwidth]{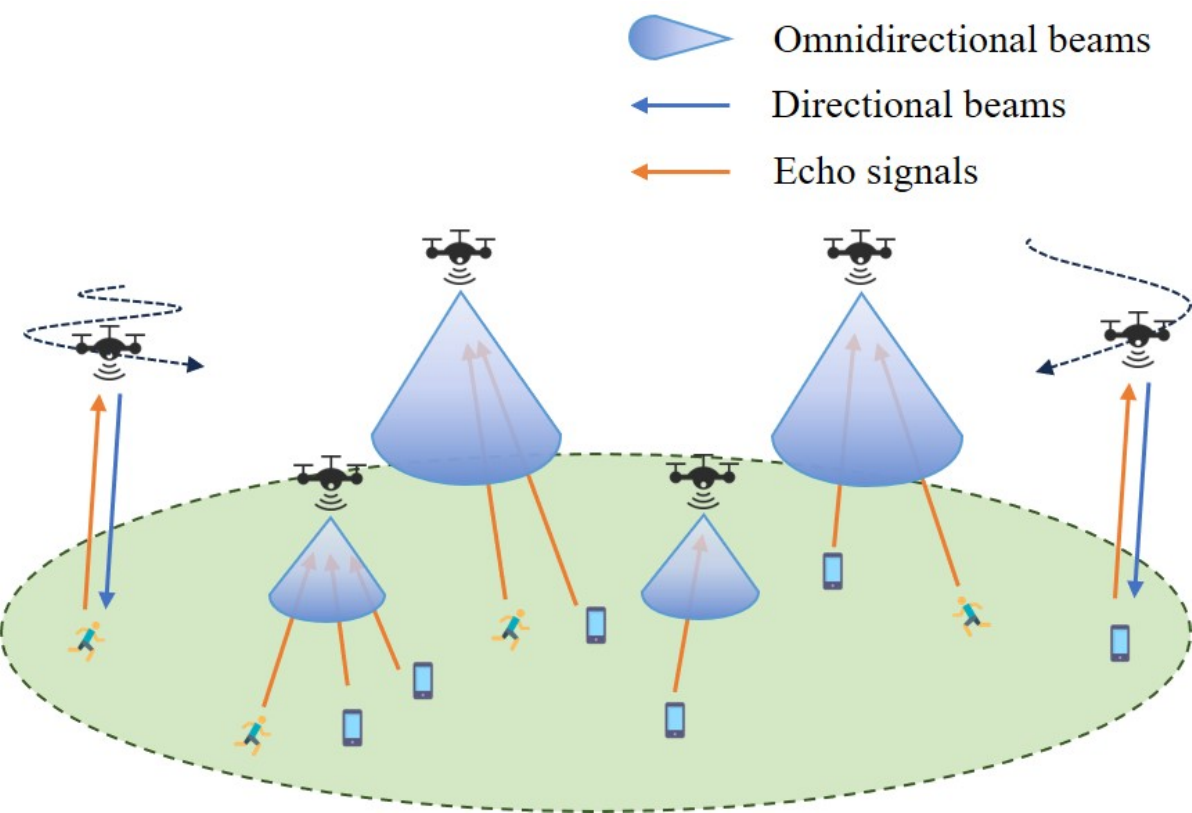}}
\caption{Multi-UAV ISAC system model.}
\label{fig.scene}
\vspace{-0.3 cm}
\end{figure}
In this section, we first introduce the application scenario of the multi-UAV network and two categories of UAVs. Subsequently, we provide an discussion of relevant models.
\subsection{Multi-UAV ISAC Network}
As shown in Fig. \ref{fig.scene}, We consider a group of UAVs is stationed at predetermined location. Their objective is to detect potential mobile users on the ground, and integrate them into the multi-UAV network for tracking and communication purposes. Note that radar can identify communication devices by the radar cross section (RCS) characteristics of the echoes. For simplicity and without loss of generality, UAVs are assumed not to interfere with each other via frequency division multiple access (FDMA). Moreover, we assume that each UAV establishes communication links with users via LoS channels and is equipped with UPA parallel to the horizontal plane. The transmit and receive arrays of the UPA both have $M_x$ rows and $M_y$ columns, i.e., each array's total number of antennas are respectively $M_t=M_r=M_x\times M_y$. It is assumed that the isolation between transmit and receive arrays is sufficient to effectively mitigate interference.

Specifically, we divide the UAVs into two categories. One is the statical UAVs, which are statically deployed within the area to ensure coverage of the entire region. These UAVs periodically sense any users entering the area and maintain communication and tracking. The other category is the mobile UAVs, which are sent to the appropriate place to provide services. In order to ensure service in the emergency scenario, we assume that UAVs are equipped with large batteries, so the energy consumption of their movement is not considered in this work, and energy constraints will be addressed in future research.
 
We divide an ISAC frame into multiple slots. In the first slot, statical UAVs are assumed initially have vague sensing data about some users, while having no information about the rest users. Therefore, the omnidirectional ISAC signal transmitted by UAVs can perform communication for users with partial information, but can perform only sensing for the users without any information at all. Based on collected sensing information, all UAVs share information among each other to distribute respective loaded users. Mobile UAVs join the multi-UAV network in the second slot. Subsequently, all UAVs collaboratively optimizes directional ISAC beams to enhance sensing and communication. Particularly, mobile UAVs can move freely within the network range and to areas with a high density of users. The iterative process of sharing information and optimizing beams continues until the end of an ISAC frame. The network then repeats this cycle in subsequent frames.

To clarify the load users by UAVs mentioned above, we define a 
 $U\times K$ indicator  matrix $\textbf{P}_n$, where the numbers of UAVs and users are $U$ and $K$, repectively. $U$ UAVs form the set $\mathcal{U}=\{1,2,\dots ,U\}$, and $K$ users form the set $\mathcal{K}=\{1,2,\dots,K\}$. The element of $\textbf{P}_n$ at the $u$-th row and $k$-th column of an array is represented as $\rho_{u,k,n}$, which indicates whether UAV $u$ is loaded with user $k$ in slot $n$:
\setlength{\arraycolsep}{1pt}
\begin{align}
{\rho _{u,k,n}} = \begin{cases}
{1,~{\rm{ if~user~}}k{\rm{~is~loaded~by~UAV~}}u{\rm{~in~slot~}}n{\rm{;}}}\\
{0,~{\rm{ otherwise}}{\rm{.}}}
\end{cases}
\end{align}
To ensure the quality of service for each user, we assume each UAV can carry a maximum of $K_{\text{max}}$ users, while each user can connect to at most one UAV.
\subsubsection{Communication Model}
With the above definition, the downlink transmission ISAC streams of UAV $u$ at time $t$ in slot $n$ can be expressed as
\begin{align}
{{\bf{s}}_{u,n}}(t) = {[{\rho _{u,1,n}} \cdot {s_{u,1,n}}(t),...,{\rho _{u,K,n}}\cdot {s_{u,K,n}}(t)]^T},
\end{align}
where the $k$-th signal ${\rho _{u,k,n}} \cdot s_{u,k,n}(t)$ carries the information to user $k$, with independent circularly symmetric complex Gaussian (CSCG) random variables with zero mean and unit variance, represented by $s_{u,k,n}(t)\sim\mathcal{CN}(0,1)$. If $\rho _{u,k,n}=0$, UAV $u$ sends no information to user $k$. Further, the trasmitted signal can be denoted by
\begin{align}
{{\bf{S}}_{u,n}}(t) = {{\bf{W}}_{u,n}}{{\bf{s}}_{u,n}}(t),
\end{align}
where ${{\bf{W}}_{u,n}} = \left[ {{{\bf{w}}_{u,1,n}}, \cdots ,{{\bf{w}}_{u,K,n}}} \right] \in {\mathbb{C}^{{M_t} \times K}}$ is the beamforming matrix. The channel from UAV $u$ to user $k$ is represented by
\begin{align}
{{\bf{h}}_{u,k,n}} = \frac{{\sqrt \alpha  }}{{{d_{u,k,n}}}}{e^{j\frac{{2\pi }}{\lambda }{d_{u,k,n}}}}\boldsymbol{a}({\theta _{u,k,n}},{\varphi _{u,k,n}}),
\end{align}
where $\alpha$ denotes the path loss at the reference distance $d_0=1$m, following \cite{10061429}:
\begin{align}
    \alpha \left( {{\rm{dB}}} \right) = 32.4 + 20\lg {f_c}\left( {{\rm{MHz}}} \right) + \left( {20 \times \eta } \right)\lg d\left( {{\rm{km}}} \right),
\end{align}
with $\eta=1$ being the path loss factor corresponding to the carrier frequency  and electromagnetic propagation environment. Besides, ${d_{u,k,n}} = \left\| {{\bf{L}}_{u,n}^{\text{UAV}} - {{\bf{L}}_{k,n}}} \right\|$ is the distance between UAV $u$ and user $k$, where $\textbf{L}_{u,n}^{\text{UAV}} = \left( {x_{u,n}^{\text{UAV}},y_{u,n}^{\text{UAV}},{h}} \right)$ and ${\textbf{L}_{k,n}} = \left( {{x_{k,n}},{y_{k,n}},0} \right)$  indicate the Cartesian coordinates of the $u$-th UAV and the $k$-th user, respectively. In particular, the steering vector $\boldsymbol{a}({\theta _{u,k,n}},{\varphi _{u,k,n}})$ of the transmit array is given in (\ref{steering}), where $\lambda$ and $\Delta d$ denote the carrier wavelength and antenna spacing of the UPA. $m_x$ and $m_y$ index the rows and columns of the UPA, respectively. ${\theta _{u,k,n}}$ and ${\varphi _{u,k,n}}$ are the elevation and azimuth angles between UAV $u$ and user $k$ in slot $n$, respectively.
\begin{table*}[b]
\normalsize
\hrulefill % 替代\hline，在公式顶部画横线
\begin{equation}\label{steering}
\begin{aligned}
&\boldsymbol{a}({\theta _{u,k,n}},{\varphi _{u,k,n}}) 
= \left( 1, \cdots ,{e^{ - j2\pi \frac{{\Delta d}}{\lambda }( {{m_x} - 1} )\sin ( {{\theta _{u,k,n}}} )\cos ( {{\varphi _{u,k,n}}} )}}, \cdots ,{e^{ - j2\pi \frac{{\Delta d}}{\lambda }( {{M_x} - 1} )\sin ( {{\theta _{u,k,n}}} )\cos ( {{\varphi _{u,k,n}}} )}} \right)\\[1ex]
& \quad\otimes \left( 1, \cdots ,{e^{ - j2\pi \frac{{\Delta d}}{\lambda }( {{m_y} - 1} )\sin ( {{\theta _{u,k,n}}} )\sin ( {{\varphi _{u,k,n}}} )}}, \cdots ,{e^{ - j2\pi \frac{{\Delta d}}{\lambda }( {{M_y} - 1} )\sin ( {{\theta _{u,k,n}}} )\sin ( {{\varphi _{u,k,n}}} )}} \right)
\end{aligned}
\end{equation}
\end{table*}

The received signal of user $k$ can be expressed as
\begin{align}
{c_{k,n}}(t) = {\bf{h}}_{u,k,n}^H{e^{j2\pi {\omega _{u,k,n}}t}}{\textbf{S}_{u,n}}(t) + {z_C}(t),
\end{align}
with $z_C(t)\sim\mathcal{CN}(0,\sigma_\mathrm{C}^2)$ symbolizing the additive white Gaussian noise (AWGN) at the receiver. $\omega_{u,k,n}$ represents the Doppler frequency. For omnidirectional transmit signal, the received signal-to-interference-plus-noise ratio (SINR) of the $k$-th user is represented as
\begin{align}
\gamma_{u,k,n}=\frac{\alpha_0^2{p_{u,n}}d_{u,k,n}^{-2}K_u^{-1}}{\alpha_0^2{p_{u,n}}(K_u-1)d_{u,k,n}^{-2}K_u^{-1}+\sigma_\mathrm{C}^2},
\end{align}
where $K_u$ indicates the numbers of users loaded by UAV $u$. The term ${p_{u,n}}$ denotes the transmit power of UAV $u$ at slot $n$. For directional transmit signal, the received SINR of the $k$-th user is represented as
\begin{align}
{\gamma _{u,k,n}} = \frac{{|{\bf{h}}_{u,k,n}^H{{\bf{w}}_{u,k,n}}{|^2}}}{{\sum\limits_{i = 1,i \ne k}^{{K}} | {\bf{h}}_{u,k,n}^H{{\bf{w}}_{u,i,n}}{|^2} + \sigma _\mathrm{C}^2}},
\end{align}
with $\sum\nolimits_{k = 1}^K {{{\left\| {{{\bf{w}}_{u,k,n}}} \right\|}^2}}  = {p_{u,n}}$. Consequently, the achievable rate of UAV $u$ at slot $n$ is formulated as
\begin{align}
{R_{u,n}} = \sum\limits_{k = 1}^{{K}} {{{\log }_2}(1 + {\gamma _{u,k,n}})}. 
\end{align}
The difference between statical UAVs and mobile UAVs leads to different evolution models and sensing models, which are discussed in turn.
\subsection{Statical UAVs}
Next, we focus on the derivation of models related to statical UAVs.
\subsubsection{Evolution Model}
\begin{figure}[h] %htbp
\centerline{\includegraphics[width=0.4\textwidth]{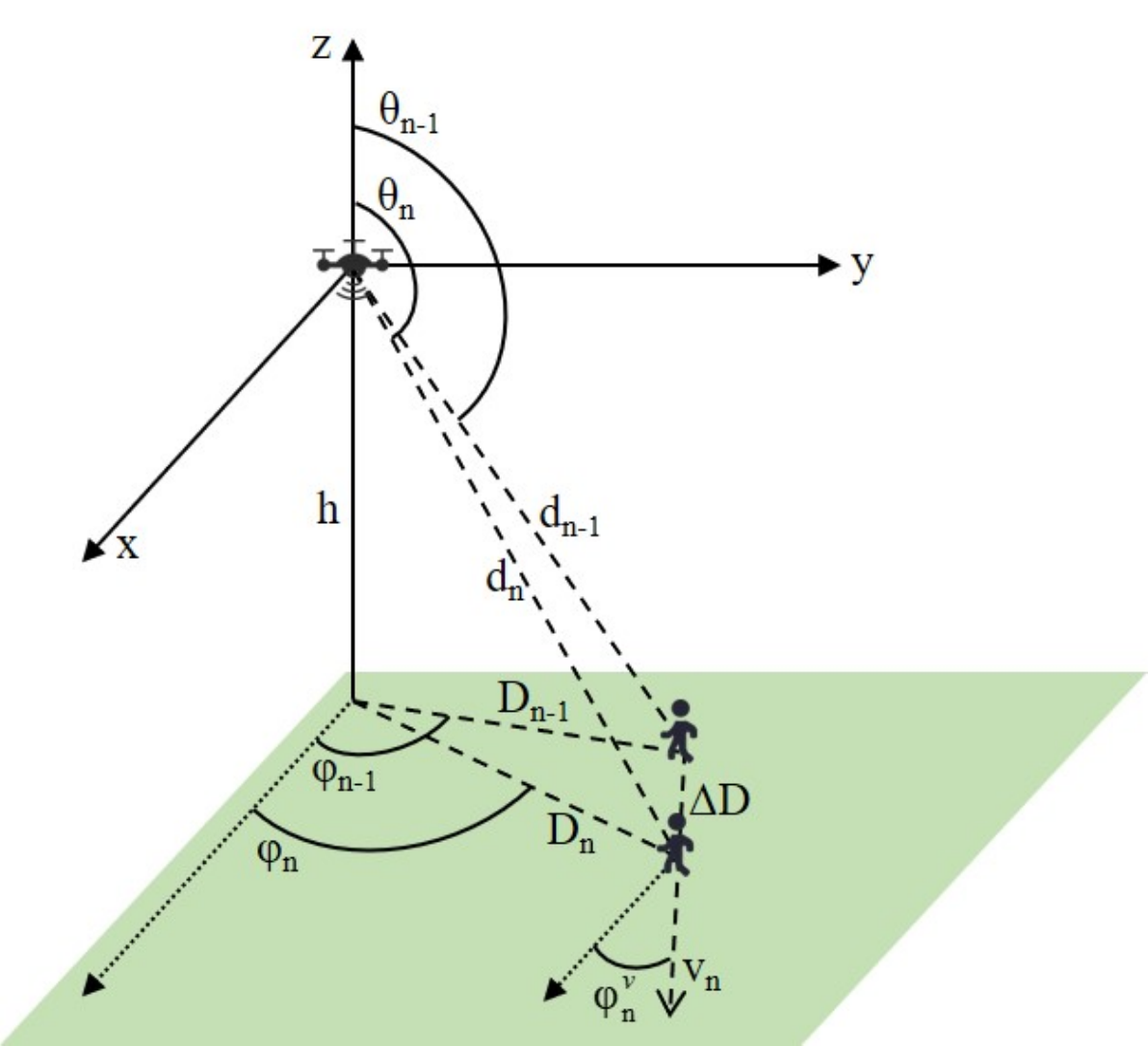}}
\caption{Evolution model.}
\label{fig.evolution}
\vspace{-0.3 cm}
\end{figure}
To obtain high precision tracking, we predict the motion parameters of users. For notational convenience, we focus on the tracking of one user by a single UAV. The elevation angle, azimuth angle, distance of the user relative to the UAV, as well as its own speed and travel direction can be expressed as: $\theta_n$, $\varphi_n$, $d_n$, $v_n$ and $\varphi^v_n$, respectively. In addition, the height of all UAVs is uniformly set as $h$. A Cartesian coordinate system is thus established as shown in Fig. \ref{fig.evolution}.

Assuming that the speed and the direction of motion hardly change in $\Delta T$, we have $v_n \approx v_{n-1}$ and $\varphi_n^v \approx \varphi_{n-1}^v$. Omitting detailed derivations, we propose the evolution model as follows
% Hence, we propose the evolution model as follows
\begin{align}\label{evolution}
\left\{\begin{array}{l}
{\theta _n} = {\theta _{n - 1}} + {v_{n - 1}}\Delta T\cos \left( \varphi^\Delta_{n-1} \right)\cos \left( {{\theta _{n - 1}}} \right)d_{n - 1}^{ - 1} + {\eta _\theta },\\
{\varphi_n} = {\varphi _{n - 1}} - {v_{n - 1}}\Delta T\sin \left( \varphi^\Delta_{n-1} \right)\csc \left( {{\theta _{n - 1}}} \right)d_{n - 1}^{ - 1} + {\eta _\varphi },\\
{d_n} = {d_{n - 1}} + {v_{n - 1}}\Delta T\cos \left( \varphi^\Delta_{n-1} \right){\rm{sin}}\left( {{\theta _{n - 1}}} \right) + {\eta _d},\\
{v_n} = {v_{n - 1}} + {\eta _v},\\
\varphi _n^v = \varphi _{n - 1}^v + {\eta _{{\varphi ^v}}},
\end{array} \right.
\end{align}
with $\varphi^\Delta_{n-1}={{\varphi _{n - 1}} - \varphi _{n - 1}^v}$. To clarify, $\eta_\theta$, $\eta_\varphi$, $\eta_d$, $\eta_v$ and $\eta _{{\varphi ^v}}$ denote the corresponding noises, which stemmed from approximation and other systematic errors.
\subsubsection{Sensing Model}
After the reflection of the ISAC signal, UAV $u$ received an echo signal, which is given in the form
\begin{align}
{{\bf{r}}_{u,k,n}}(t) &= {g_{u,k,n}}{e^{j2\pi {\mu _{u,k,n}}t}}{\bf{b}}({\theta _{u,k,n}},{\varphi _{u,k,n}})\nonumber \\ 
&\cdot{\boldsymbol{a}^H}({\theta _{u,k,n}},{\varphi _{u,k,n}}) {{\bf{S}}_{u,n}}(t - {\tau _{u,k,n}}) + {{\bf{z}}_r}(t),
\end{align}
where ${\bf{b}}({\theta _{u,k,n}},{\varphi _{u,k,n}})$ is the steering vevtor of the receive array similar to ${\boldsymbol{a}}({\theta _{u,k,n}},{\varphi _{u,k,n}})$, and ${{\bf{z}}_r}(t)$ represents the zero-mean complex AWGN with variance $\sigma^2$. The terms $\mu _{u,k,n}$ and $\tau _{u,k,n}$ are the Doppler frequency and time delay for the $k$-th user during the $n$-th slot. According to  \cite{10529184}, the reflection coefficient is expressed as ${g_{u,k,n}} = \sqrt \beta{d_{u,k,n}^{-2}}$, 
and based on Friis free space formula, we have $\beta={{G_\mathrm{T}}{G_\mathrm{R}}{\lambda ^2}\varepsilon }{{\left( {4\pi } \right)}^{-3}}$, where ${G_\mathrm{T}}$ and ${G_\mathrm{R}}$ are the transmit and receive antenna gains, respectively. It is assumed that RCS $\varepsilon$ is the average of its distribution model \cite{10.1049/iet-rsn.2018.5646} to simplify analysis. For omnidirectional echo signal, the received SNR from the $k$-th user is calculated as $\mathrm{SNR}_{{u,k,n}}^0 = {{{p_{u,n}}}|{g_{u,k,n}}{|^2}}{{\sigma ^{-2}}}$.
For directional echo signal, the received SNR of the $k$-th user is calculated as
\begin{align}
\mathrm{SNR}_{{u,k,n}}^0 = \frac{{|{g_{u,k,n}}{|^2}\sum\limits_{i = 1}^K {{{\left| {{\boldsymbol{a}^H}({\theta _{u,k,n}},{\varphi _{u,k,n}}){{\bf{w}}_{u,i,n}}} \right|}^2}} }}{{{\sigma ^2}}},
\end{align}
where $\sum\limits_{i = 1}^K {{{\left| {{\boldsymbol{a}^H}({\theta _{u,k,n}},{\varphi _{u,k,n}}){{\bf{w}}_{u,i,n}}} \right|}^2}}$ denotes beam gain for user $k$.

Next, as mentioned in \cite{10061429}, we can employ the classical Multiple Signal classification (MUSIC) algorithm to estimate the elevation and azimuth, then the measurement model of angles takes the following  form
\begin{equation}
\begin{aligned}
{\hat \theta _{u,k,n}} &= {\theta _{u,k,n}} + {z_{{\theta _{u,k,n}}}}, \\
{\hat \varphi _{u,k,n}} &= {\varphi _{u,k,n}} + {z_{{\varphi _{u,k,n}}}},\label{estiPhi}
\end{aligned}
\end{equation}
where ${z_{{\theta _{u,k,n}}}}$ and ${z_{{\varphi _{u,k,n}}}}$ signify Gaussian measurement noises with zero mean and variances $\sigma^2_{\theta_{u,k,n}}$ and $\sigma^2_{\varphi_{u,k,n}}$, respectively. 
After that, the received beamformer is constructed based on the estimated elevation and azimuth. The weighted echo signal thus can be formulated in (\ref{weighted}).
\begin{figure*}[!b]
\normalsize
\hrulefill
\begin{equation}\label{weighted}
\begin{aligned}
{r_{u,k,n}}(t) 
= {g_{u,k,n}} e^{j2\pi {\mu _{u,k,n}} t}\sqrt{\frac{1}{M_r}}
{\bf{b}}^H(\hat{\theta}_{u,k,n}, \hat{\varphi}_{u,k,n})
{\bf{b}}(\theta_{u,k,n}, \varphi_{u,k,n}) 
{\boldsymbol{a}}^H(\theta_{u,k,n}, \varphi_{u,k,n})
{\bf{S}}_{u,n}(t - \tau_{u,k,n}) + z_r(t)
\end{aligned}
\end{equation}
%\vspace*{4pt}
\end{figure*}
% \begin{figure*}[tbp]
% \begin{align}
% \begin{array}{l}\label{weighted}
% {r_{u,k,n}}(t) = {g_{u,k,n}}{e^{j2\pi {\mu _{u,k,n}}t}}\sqrt {\frac{1}{{{M_r}}}} {{\bf{b}}^H}({\hat \theta _{u,k,n}},{\hat \varphi _{u,k,n}}){\bf{b}}({\theta _{u,k,n}},{\varphi _{u,k,n}}) {\boldsymbol{a}^H}({\theta _{u,k,n}},{\varphi _{u,k,n}}){{\bf{S}}_{u,n}}(t - {\tau _{u,k,n}}) + {z_r}(t)
% \end{array}
% \end{align}
% \end{figure*}
Hence, the received SNR from user $k$ is rewritten as
\begin{align}
\mathrm{SNR}_{{u,k,n}}^1 &= \mathrm{SNR}_{{u,k,n}}^0\cdot \nonumber \\
 &\frac{1}{{{M_r}}}{\left| {{{\bf{b}}^H}({{\hat \theta }_{u,k,n}},{{\hat \varphi }_{u,k,n}}){\bf{b}}({\theta _{u,k,n}},{\varphi _{u,k,n}})} \right|^2}.
\end{align}
Subsequently, upon the application of matched filtering (MF), the measurement model for ascertaining the time delay $\tau_{u,k,n}$ and Doppler shift $\mu_{u,k,n}$ is given in the form:
\begin{equation}
\begin{aligned}
{\hat \tau _{u,k,n}} &= \frac{{2{d_{u,k,n}}}}{c} + {z_{{\tau _{u,k,n}}}}, \\
{\hat \mu _{u,k,n}} &= \frac{{2{v_{k,n}}\cos \varphi^\Delta_{n-1}\sin \left( {{\theta _{u,k,n}}} \right)}}{\lambda } + {z_{{\mu _{u,k,n}}}},\label{estiMu}
\end{aligned}
\end{equation}
where that ${z_{{\tau _{u,k,n}}}}$ and ${z_{{\mu _{u,k,n}}}}$ signify Gaussian measurement noises with zero mean and variances $\sigma^2_{\tau_{u,k,n}}$ and $\sigma^2_{\mu_{u,k,n}}$, respectively. 

To better illustrate, we derive the variances of ${z_{{\theta _{u,k,n}}}}$, ${z_{{\varphi _{u,k,n}}}}$, ${z_{{\tau _{u,k,n}}}}$ and ${z_{{\mu _{u,k,n}}}}$. In fact, we apply the MUSIC algorithm to estimate the spatial frequencies related to (\ref{steering})
\begin{align}
a_{\mathrm{sf}} = \pi \sin \theta \cos \varphi,~~~~b_{\mathrm{sf}} = \pi \sin \theta \sin \varphi,
\end{align}
where the subscripts of $u,k,n$ are omitted for the rest of this sub-section for convenience. Since a closed-form solution for the actual variance is arduous to obtain, the Cram$\acute{\text{e}}$r-Rao Lower Bound (CRLB) is often used as an alternative \cite{10.5555/151045}. The CRLBs of $a_{\mathrm{sf}}$ and $b_{\mathrm{sf}}$ are expressed as \cite{10061429}
\begin{align}
{C_{a_{\mathrm{sf}}}} = {C_{b_{\mathrm{sf}}}} = \frac{6}{{N_{\text{sam}}^2{M_r}\mathrm{SNR}^0}},
\end{align}
where $N_{\text{sam}}$ is the number of signal samples. A simple conversion of spatial frequencies yield elevation and azimuth 
\begin{align}\label{eleazi}
\theta  = \arcsin \left( {\frac{{\sqrt {{a_{\mathrm{sf}}^2} + {b_{\mathrm{sf}}^2}} }}{\pi }} \right),~~~~ \varphi  = \arctan \left( {\frac{b_{\mathrm{sf}}}{a_{\mathrm{sf}}}} \right).
\end{align}
With the empirical measurement errors of the MUSIC algorithm following the Gaussian distribution \cite{10061429}, we linearize (\ref{eleazi}). The covariance matrix can accordingly be expressed as
\begin{align}
{\boldsymbol{\sigma}} _{{\theta ,\varphi }}^2 = {\bf{CRL}}{{\bf{B}}_{\theta ,\varphi }} = \left[ {\begin{array}{*{20}{c}}
{\frac{{\partial \theta }}{{\partial a_{\mathrm{sf}}}}}{\frac{{\partial \theta }}{{\partial b_{\mathrm{sf}}}}}\\
{\frac{{\partial \varphi }}{{\partial a_{\mathrm{sf}}}}}{\frac{{\partial \varphi }}{{\partial b_{\mathrm{sf}}}}}
\end{array}} \right]\left[ {\begin{array}{*{20}{c}}
{{C_a}}&0\\
0&{{C_b}}
\end{array}} \right]{\left[ {\begin{array}{*{20}{c}}
{\frac{{\partial \theta }}{{\partial a}}}{\frac{{\partial \theta }}{{\partial b}}}\\
{\frac{{\partial \varphi }}{{\partial a}}}{\frac{{\partial \varphi }}{{\partial b}}}
\end{array}} \right]^T},
\end{align}
and then we have
\begin{align}
\sigma _\theta ^2 = {\left[ {{\bf{CRL}}{{\bf{B}}_{\theta ,\varphi }}} \right]_{1,1}}, ~~~~\sigma _\varphi ^2 = {\left[ {{\bf{CRL}}{{\bf{B}}_{\theta ,\varphi }}} \right]_{2,2}}.
\end{align}
The CRLB of $\tau$ and $\mu$ can be obtained by MF algorithm, while the errors are not Gaussian distributed. Even so, it is still rational to treat these errors as Gaussian distributed\cite{10061429}. Accordingly, their variances are formulated as 
\begin{align}
\sigma _{{\tau }}^2 = \mathrm{CRLB}_\tau  &= \frac{3}{{2{\pi ^2}{B_\mathrm{bw}}\chi }},\\
\sigma _{\mu }^2 = \mathrm{CRLB}_\mu  &= \frac{1}{{{{\left( {2\pi } \right)}^2}\Delta {T^2}{f_\mathrm{s}}\chi }}.
\end{align}
Here, $B_\mathrm{bw}$ and ${f_\mathrm{s}}$ represent bandwidth of the transmitted signal and the sample rate of the analog-digital converter (ADC). The output SNR of the MF is defined by $\chi  = \mathrm{SNR}^1\Delta T{B_\mathrm{bw}}$.
\subsubsection{Extended Kalman Filtering}
 Given the nonlinearity of the derived models, we employ an EKF method that utilizes the first-order Taylor expansion to approximate them. 

 Multi-UAV network utilizes measurements to substitute initial estimates at the first slot, which is
    ${\left[ {{{\bar \theta }_{u,k,1}},{{\bar \varphi }_{u,k, 1}},{{\bar d}_{u,k,1}}} \right]^T}$.
By converting the estimated angles and distance to Cartesian coordinates, we have
\begin{align}
\left\{ {\begin{array}{*{20}{c}}
{{{\bar x}_{k,1}} = {{\bar d}_{u,k,1}}\sin \left( {{{\bar \theta }_{u,k,1}}} \right)\cos \left( {{{\bar \varphi }_{u,k,1}}} \right)},\\
{{{\bar y}_{k,1}} = {{\bar d}_{u,k,1}}\sin \left( {{{\bar \theta }_{u,k,1}}} \right)\sin \left( {{{\bar \varphi }_{u,k,1}}} \right)}.
\end{array}} \right.
\end{align}
Similarly, we obtain ${{\bar x}_{k,2}}$ and ${{\bar y}_{k,2}}$ of user $k$ at slot $2$, and the direction of motion is calculated by
\begin{align}
\bar \varphi _{u,k,2}^v = \arctan \left( {\frac{{{{\bar y}_{k,2}} - {{\bar y}_{k,1}}}}{{{{\bar x}_{k,2}} - {{\bar x}_{k,1}}}}} \right).
\end{align} 

To summarize, the evolution model and the measurement model can be respectively expressed as $\mathbf{e}_n=\mathbf{g}(\mathbf{e}_{n-1}) +  \boldsymbol{\eta}_n$ and $\mathbf{m}_n=\mathbf{h}(\mathbf{e}_n) + \mathbf{z}_n$, where ${\mathbf{e}_n} = {\left[ {{\theta _n},{\varphi _n},{d_n},{v_n},\varphi _n^v} \right]^T}$, $\mathbf{m}_n = [\hat \theta_n,\hat \varphi_n,\hat \tau_n,\hat \mu_n]^T$, $\boldsymbol{\eta} = [\eta_\theta,\eta_\varphi,\eta_d,\eta_v,\eta_{\varphi^v}]^T$ and $\mathbf{z} = [z_\theta,z_\varphi,z_\tau,z_\mu]^T$. $\mathbf{g}(\cdot)$ is defined in (\ref{evolution}), and $\mathbf{h}(\cdot)$ is defined in (\ref{estiPhi}) and (\ref{estiMu}). As discussed above, both $\boldsymbol{\eta}$ and $\mathbf{z}$ can be modeled as  zero-mean Gaussian distribution with covariance matrices
\begin{align}
    \mathbf{Q}_s&=\text{diag}(\sigma_{\tilde\theta}^2,\sigma_{\tilde\varphi}^2,\sigma_{\tilde d}^2, \sigma_{\tilde v}^2, \sigma_{\tilde\varphi^v}^2),\\
    \mathbf{Q}_m&=\text{diag}({\boldsymbol{\sigma}} _{_{\theta ,\varphi }}^2,\sigma^2_\tau,\sigma^2_\mu),
\end{align}
where $\sigma_{\tilde\theta}^2,\sigma_{\tilde\varphi}^2,\sigma_{\tilde d}^2, \sigma_{\tilde v}^2$ and $\sigma_{\tilde\varphi^v}^2$ represent the variance of motion parameters. We can execute EKF procedure accordingly, please see \cite{9171304} for more details.

\subsubsection{Beam Coverage Control}
In traditional beam design, a beamforming vector is constructed as ${\bf{w}} = \boldsymbol{a}(\hat \theta ,\hat \varphi )$ based on predicted angles \cite{9557830}. Nevertheless, such beams are prone to misalignment due to inaccuracies in prediction, resulting in lower beam gain and degraded communication and sensing performance. Therefore, we strive to assess the prediction uncertainty  and calculate a beam coverage within a given confidence coefficient to maximize beam gain. It should be highlighted that prediction accuracy is affected by both measurement noise and system noise.

The predicted motion parameters follow the distribution
${\bar {\bf{e}}_{\left. n \right|n - 1}} \sim {\cal N}\left( {{{\bf{e}}_n},{\bf{MSE}}_{\left. n \right|n - 1}} \right)$, where the covariance matrix of prediction is articulated as
\begin{align}
\textbf{MSE}_{\left. n \right|n - 1} = \left( {\begin{array}{*{20}{c}}
{{\Sigma _{\theta \theta }}}&{{\Sigma _{\theta \varphi }}}& \cdots \\
{{\Sigma _{\theta \varphi }}}&{{\Sigma _{\varphi \varphi }}}& \cdots \\
 \vdots & \vdots & \ddots 
\end{array}} \right),
\end{align}
where the covariance $\text{Cov}(\theta,\varphi)$ is abbreviated as $\Sigma _{\theta \varphi }$. According to the property of marginal Gaussian distribution, the angular error between the predicted and actual values can be modeled as
\begin{align}
{\boldsymbol{\Psi} _n}   = \left[ \begin{array}{l}
\Delta {\theta _n}\\
\Delta {\varphi _n}
\end{array} \right] = \left[ \begin{array}{l}
{{\bar \theta }_{\left. n \right|n - 1}} - {\theta _n}\\
{{\bar \varphi }_{\left. n \right|n - 1}} - {\varphi _n}
\end{array} \right] \sim {\cal N}\left( {0,\bf{\Sigma}^\mathrm{rad} } \right),
\end{align}
with $\bf{\Sigma}^\mathrm{rad}$ is composed of the first two elements of the first two rows of $\textbf{MSE}_{\left. n \right|n - 1}$.
% \begin{align}
% \bf{\Sigma}^\mathrm{rad}  = \left( {\begin{array}{*{20}{c}}
% {{\Sigma _{\theta \theta }}}&{{\Sigma _{\theta \varphi }}}\\
% {{\Sigma _{\theta \varphi }}}&{{\Sigma _{\varphi \varphi }}}
% \end{array}} \right).
% \end{align}
The prediction uncertainty is portrayed by binary Gaussian distribution vector $\boldsymbol{\Psi} _n$, with the confidence interval being ellipse, as will be discussed in Section V. As a result, we take the confidence ellipse corresponding to the given confidence level 0.99 as the beam coverage. When the major and minor axes of the confidence ellipse are $2\sqrt {9.21{\lambda _1}} $ and $2\sqrt {9.21{\lambda _2}} $, respectively, from the cumulative Chi-square distribution, the confidence is counted as 0.99. Note that $\lambda_1$ and $\lambda_2$ represent the eigenvalues of $\bf{\Sigma}^\mathrm{rad}$, respectively. The angle of the elliptic direction is denoted by $\phi  = \arctan \left( {\frac{{{\boldsymbol{\nu} _1}\left( \varphi  \right)}}{{{\boldsymbol{\nu} _1}\left( \theta  \right)}}} \right)$, where $\boldsymbol{\nu} _1$ is the eigenvector of $\bf{\Sigma}^\mathrm{rad}$ corresponding to the largest eigenvalue. In short, beam coverage is controlled by the predicted angles of each slot.
\subsection{Mobile UAVs}
The greatest difference between mobile UAVs and statical UAVs lies in their speeds. For analytical simplicity, mobile UAVs are assumed to maintain a constant average flight speed within a slot duration $\Delta T$, neglecting the modeling of acceleration. Consequently, the task of utilizing mobility to alleviate multi-UAV network load is reduced to direction planning.
\subsubsection{Evolution Model}
Due to mobility, it is intricate  to incorporate the motion parameters into the aforementioned spherical coordinate system. We thus define $x_n^\mathrm{U} = {x_n} - x_n^{\text{UAV}}$, $y_n^\mathrm{U} = {y_n} - y_n^{\text{UAV}}$ and $z_n^\mathrm{U}={z_n}-h$ as the Cartesian coordinate of the user $(x_n,y_n,z_n)$ relative to the UAV $(x_n^{\text{UAV}},y_n^{\text{UAV}},h)$, and the evolution model of the motion parameters are reformulated as
\begin{align}
\left\{ \begin{array}{l}
x_n^\mathrm{U} = x_{n - 1}^\mathrm{U} + {v_{n - 1}}\Delta T\cos \varphi _{n - 1}^v\\
 ~~~~~~~- v_{n - 1}^\mathrm{U}\Delta T\cos \varphi _{n - 1}^\mathrm{U} + {\eta _{{x^\mathrm{U}}}},\\
y_n^\mathrm{U} = y_{n - 1}^\mathrm{U} + {v_{n - 1}}\Delta T\sin \varphi _{n - 1}^v\\
 ~~~~~~~- v_{n - 1}^\mathrm{U}\Delta T\sin \varphi _{n - 1}^\mathrm{U} + {\eta _{{y^\mathrm{U}}}},\\
z_n^\mathrm{U} = z_{n - 1}^\mathrm{U} + {\eta _{{z^\mathrm{U}}}},\\
{v_n} = {v_{n - 1}} + {\eta _v},\\
\varphi _n^v = \varphi _{n - 1}^v + {\eta _{\varphi^v} },
\end{array} \right.
\end{align}
where $v_{n}^\mathrm{U}$ and $\varphi _{n}^\mathrm{U}$ represent the speed and direction of mobile UAVs. Likewise, $\eta _{{x^\mathrm{U}}}$, $\eta _{{y^\mathrm{U}}}$ and $\eta _{{z^\mathrm{U}}}$ denote the corresponding noises derived from approximation and other systematic errors, with variances of $\sigma_{\tilde x^\mathrm{U}}^2,\sigma_{\tilde y^\mathrm{U}}^2$ and $\sigma_{\tilde z^\mathrm{U}}^2$.
\subsubsection{Sensing Model}
The sensing model deforms with the above evolution model. By omitting the derivation, we conclude as follows
\begin{align}
\left\{ \begin{array}{l}
{{\hat \theta }_n} = \arctan \left( {\frac{{\sqrt {{{\left( {x_n^\mathrm{U}} \right)}^2} + {{\left( {y_n^\mathrm{U}} \right)}^2}} }}{{z_n^\mathrm{U}}}} \right) + {z_{{\theta _n}}},\\
{{\hat \varphi }_n} = \arctan \left( {\frac{{y_n^\mathrm{U}}}{{x_n^\mathrm{U}}}} \right) + {z_{{\varphi _n}}},\\
{{\hat \tau }_n} = \frac{{2d_n^\mathrm{U}}}{c} + {z_{{\tau _n}}},\\
{{\hat \mu }_n} = \frac{{2{v_n}}}{\lambda }\left( {\frac{{x_n^\mathrm{U}}}{{d_n^\mathrm{U}}}\cos \left( {\varphi _{_n}^v} \right) + \frac{{y_n^\mathrm{U}}}{{d_n^\mathrm{U}}}\sin \left( {\varphi _{_n}^v} \right)} \right) + {z_{{\mu _n}}},
\end{array} \right.
\end{align}
where $d_n^\mathrm{U} = \sqrt {{{\left( {x_n^\mathrm{U}} \right)}^2} + {{\left( {y_n^\mathrm{U}} \right)}^2} + {{\left( {z_n^\mathrm{U}} \right)}^2}} $. Accordingly, the similar EKF procedure can be applied. %like (\ref{ekfbegin})-(\ref{ekfend}).
\subsubsection{Beam Coverage Control}
To obtain the estimation error of the predicted angles, we linearly approximate the reconstructed motion parameters. The covariance matrix of prediction is rewritten as
\begin{align}
\textbf{MSE}_{\left. n \right|n - 1} = \left( {\begin{array}{*{20}{c}}
{{\Sigma _{xx}}}&{{\Sigma _{xy}}}&{{\Sigma _{xz}}}& \cdots \\
{{\Sigma _{xy}}}&{{\Sigma _{yy}}}&{{\Sigma _{yz}}}& \cdots \\
{{\Sigma _{xz}}}&{{\Sigma _{yz}}}&{{\Sigma _{zz}}}& \cdots \\
 \vdots & \vdots & \vdots & \ddots 
\end{array}} \right).
\end{align}
With the distribution of predicted parameters ${\bar {\bf{e}}_{\left. n \right|n - 1}}$ in hand, one can rederive of ${\boldsymbol{\Psi} _n}$ as
\begin{align}
{\boldsymbol{\Psi} _n} =
\left[ {\begin{array}{*{20}{c}}
{\Delta {\theta _n}}\\
{\Delta {\varphi _n}}
\end{array}} \right] \approx \left[ {\begin{array}{*{20}{c}}
{\begin{array}{*{20}{c}}
{\frac{{\partial \theta }}{{\partial x}}}{\frac{{\partial \theta }}{{\partial y}}}{\frac{{\partial \theta }}{{\partial z}}}
\end{array}}\\
{\begin{array}{*{20}{c}}
{\frac{{\partial \varphi }}{{\partial x}}}{\frac{{\partial \varphi }}{{\partial y}}}{\frac{{\partial \varphi }}{{\partial z}}}
\end{array}}
\end{array}} \right]\left[ {\begin{array}{*{20}{c}}
{{{\bar x}_{\left. n \right|n - 1}} - {x_n}}\\
{{{\bar y}_{\left. n \right|n - 1}} - {y_n}}\\
{{{\bar z}_{\left. n \right|n - 1}} - {z_n}}
\end{array}} \right],
\end{align}
with 
\begin{align}
\left[ {\begin{array}{*{20}{c}}
{{{\bar x}_{\left. n \right|n - 1}} - {x_n}}\\
{{{\bar y}_{\left. n \right|n - 1}} - {y_n}}\\
{{{\bar z}_{\left. n \right|n - 1}} - {z_n}}
\end{array}} \right] = \left[ {\begin{array}{*{20}{c}}
{\Delta x}\\
{\Delta y}\\
{\Delta z}
\end{array}} \right] \sim {\cal N}\left( {0,\bf{\Sigma} } \right),
\end{align}
with $\bf{\Sigma}$ is composed of the first three elements of the first three rows of $\textbf{MSE}_{\left. n \right|n - 1}$. Thus we have
\begin{align}
{\boldsymbol{\Psi} _n} \!\sim\! {\cal N}\left( {0,{\bf{\Sigma} ^{\boldsymbol{\Psi}} }} \right), ~{\bf{\Sigma} ^{\boldsymbol{\Psi}} } \!=\! \left[ {\begin{array}{*{20}{c}}
{\begin{array}{*{20}{c}}
{\frac{{\partial \theta }}{{\partial x}}}{\frac{{\partial \theta }}{{\partial y}}}{\frac{{\partial \theta }}{{\partial z}}}
\end{array}}\\
{\begin{array}{*{20}{c}}
{\frac{{\partial \varphi }}{{\partial x}}}{\frac{{\partial \varphi }}{{\partial y}}}{\frac{{\partial \varphi }}{{\partial z}}}
\end{array}}
\end{array}} \right]\!\!\bf{\Sigma} {\left[ {\begin{array}{*{20}{c}}
{\begin{array}{*{20}{c}}
{\frac{{\partial \theta }}{{\partial x}}}{\frac{{\partial \theta }}{{\partial y}}}{\frac{{\partial \theta }}{{\partial z}}}
\end{array}}\\
{\begin{array}{*{20}{c}}
{\frac{{\partial \varphi }}{{\partial x}}}{\frac{{\partial \varphi }}{{\partial y}}}{\frac{{\partial \varphi }}{{\partial z}}}
\end{array}}
\end{array}} \right]^T}.
\end{align}
Finally, beam coverage can be calculated by the confidence ellipse deduced in Section II-B.
\section{Problem Formulation}
In this section,  we formulate a joint optimization problem for multi-UAV network. At the start of each slot, the multi-UAV network design beamforming matrices, payload allocations, and direction of mobile UAVs in the light of the collected sensing information, aiming to maximize the total achievable rate in that slot. To maintain generality, all UAVs and users are assumed stationary within $\Delta T$.
The predicted channel from UAV $u$ to user $k$ in slot $n$ is denoted by
\begin{align}
{{\bar {\bf{h}}}_{_{u\!,k\!,\!\left. n\! \right|\!n - 1}}} \!= \!\frac{{\sqrt \alpha  }}{{{{\bar d}_{_{u\!,k\!,\!\left. n\! \right|\!n - 1}}}}}{e^{j\!\frac{{2\pi }}{\lambda }\!{{\bar d}_{_{u\!,k\!,\!\left. n\! \right|\!n - 1}}}}} \boldsymbol{a}({{\bar \theta }_{_{u\!,k\!,\!\left. n\! \right|\!n - 1}}},{{\bar \varphi }_{_{u\!,k\!,\!\left. n\! \right|\!n - 1}}}).
\end{align}
 The achievable rate between UAV $u$ and user $k$ at slot $n$ is expressed by
\begin{align}
{\bar R\!_{_{u\!,k\!,\!\left. n\! \right|\!n \!-\! 1}}} \!\!\!=\! {\log _2}\!\!\left(\!\! {1 \!\!+\! \frac{{{\rm{tr}}({{\bar {\bf{h}}}_{_{u\!,k\!,\!\left. n\! \right|\!n \!- \!1}}}\!\bar {\bf{h}}_{_{u\!,k\!,\!\left. n\! \right|\!n \!- \!1}}^H\!\!{{\bf{W}}\!\!_{_{u\!,k\!,n}}})}}{{\sum\limits_{_{i = 1,i \ne k}}^K \!\!\!\!{{\rm{tr}}} ({{\bar {\bf{h}}}_{_{u\!,k\!,\!\left. n\! \right|\!n \!-\! 1}}}\!\bar {\bf{h}}_{_{u\!,k\!,\!\left. n\! \right|\!n \!-\! 1}}^H\!\!{{\bf{W}}\!\!_{_{u\!,i\!,n}}}) \!+\! \sigma _{_C}^2}}} \!\!\right),
\end{align}
 where ${{\bf{W}}_{u,k,n}} = {{{\bf{w}}_{u,k,n}}}{{{\bf{w}}^H_{u,k,n}}}$.
Accordingly, the aforementioned  mission of tracking and communication of multi-UAV network is refined into the optimization problem as follows
%\begin{subequations}
\begin{align}
%\begin{array}{l}
&\mathop {{\rm{max}}}\limits_{\{ {{\bf{W}}_{u,k,n}}\} ,{{\rm \textbf{P}}_n},\varphi _{u,n}^\mathrm{U}} \sum\limits_{u = 1}^U {\sum\limits_{k = 1}^K {{\rho _{u,k,n}}{{\bar R}_{u,k,\left. n \right|n - 1}}} } \label{problem}\\ 
&~~~~~~~\;{\rm{s}}{\rm{.t}}{\rm{.}}~~{p_{u,n}} \le {P_{Tmax }},\forall u,n,\tag{\ref{problem}{a}}\label{problema}\\
&~~~~~~~~~~~~~{{\bf{W}}_{u,k,n}}\succeq0,{{\bf{W}}_{u,k,n}} = {\bf{W}}_{u,k,n}^H,\forall u,k,n,\tag{\ref{problem}{b}}\label{problemb}\\
&~~~~~~~~~~~~~\;{\rm{rank}}({{\bf{W}}_{u,k,n}}) = 1,\forall u,k,n,\tag{\ref{problem}{c}}\label{problemc}\\
&~~~~~~~~~~~~~\;{{\bar R}_{u,k,\left. n \right|n - 1}} \ge {\Gamma _{u,k}},\forall u,k,n,\tag{\ref{problem}{d}}\label{problemd}\\
&~~~~~~~~~~~~~~{\rho _{u,k,n}} \in \left\{ {0,1} \right\},\forall u,k,n,\tag{\ref{problem}{e}}\label{probleme}\\
&~~~~~~~~~~~~~\sum\limits_{u = 1}^U {{\rho _{u,k,n}}}  \le 1,\forall k,n,\tag{\ref{problem}{f}}\label{problemf}\\
&~~~~~~~~~~~~~\sum\limits_{k = 1}^K {{\rho _{u,k,n}}}  \le {K_{\max }},\forall u,n,\tag{\ref{problem}{g}}\label{problemg}\\
&~~~~~~~~~~~~~{\boldsymbol{a}^H}(o){{\bf{W}}_{u,k,n}}\boldsymbol{a}(o) + {B_k}{\rm{tr}}({{\bf{W}}_{u,k,n}}) \ge \nonumber\\
&\boldsymbol{a}^H({{\bar \theta }_{_{u\!,k\!,\!\left. n\! \right|\!n - 1}}},{{\bar \varphi }_{_{u\!,k\!,\!\left. n\! \right|\!n - 1}}}){{\bf{W}}_{u,k,n}}\boldsymbol{a}({{\bar \theta }_{_{u\!,k\!,\!\left. n\! \right|\!n - 1}}},{{\bar \varphi }_{_{u\!,k\!,\!\left. n\! \right|\!n - 1}}}),\nonumber\\
&~~~~~~~~~~~~~~~~~~~~~~~~~~~~~~~~~~~~~\forall u,k,n,\forall o \in {\rm O}_{u,k,n},\tag{\ref{problem}{h}}\label{problemh}
%\end{array}
\end{align}
%\end{subequations}
where $o = (\theta ,\varphi )$, and ${\rm O}_{u,k,n}$ denotes beam coverage of UAV $u$ to user $k$ in slot $n$. 

Specifically, the objective function (\ref{problem}) is the sum of the downlink achievable rates between each UAV and its loaded users. (\ref{problema}) represents the transmit power constraint. The equations (\ref{problemb}) and (\ref{problemc}) on ${{\bf{W}}_{u,k,n}}$  require that it  must satisfy semidefinite, Hermitian, and rank-one conditions. The parameter ${\Gamma _{u,k}}$ regulates the minimum achievable rate from UAV $u$ to user $k$. (\ref{probleme}) ensures the binary nature of of the load index $\rho_{u,k,n}$, (\ref{problemf}) indicates that each user can only link one UAV, and (\ref{problemg}) indicates the maximum number of users that each UAV can load. To guarantee that the directional beams cover the users, one way is to activate a varying number of antennas to make the half-power beamwidth approach the calculated confidence ellipse \cite{10061429}. However, this method introduces errors between the beam coverage and the confidence ellipse, while the proposed constraint (\ref{problemh}) avoids these drawbacks. In (\ref{problemh}), ${\boldsymbol{a}^H}(\cdot){{\bf{W}}_{u,k,n}}\boldsymbol{a}(\cdot)$ reveals the beampattern gain of UAV $u$ for user $k$, and $B_k$  is a controllable coefficient. Ideally, we want the beampattern gain of UAV $u$ for user $k$ to be equal within the coverage of the confidence ellipse, with the predicted angle at the center, which corresponds to the ideal radar beampattern \cite{4350230}. However, such optimization conditions are too stringent and difficult to compute. Therefore, we take ${B_k}{\rm{tr}}({{\bf{W}}_{u,k,n}})$ as an acceptable gain mismatch, allowing the gain within the confidence ellipse to approximate the gain in the predicted direction.

Problem (\ref{problem}) is undoubtedly an NP-hard problem. First, the load matrix $\textbf{P}_n$, being optimization variables, can only take integer values, i.e., (\ref{probleme})-(\ref{problemg}) are integer constraints. Second, since the purpose of mobile UAVs is to alleviate load pressure, optimizing only the direction for the next slot likely fall into local optima. Finally, even if the load matrix and direction of mobile UAVs are determined, problem (\ref{problem}) remains non-convex due to the presence of constraint (\ref{problemc}). Overall, problem (\ref{problem}) is challenging to solve in closed-form. We then propose solving the original problem in three steps  to obtain a sub-optimal but effective solution.
\section{Proposed Solution}
In this section, we sequentially decompose problem (\ref{problem}) into the steps of beam optimization, load optimization, and direction planning, according to the interrelation of the optimization variables. Each of these steps will be discussed in turn. By combining the distributed computing power of each UAV, we propose an improved SCA-IRM algorithm to solve the beam optimization. For fairness, the loading optimization problem of the multi-UAV network can be solved iteratively by adopting the coalition game approach. In order to ensure global optimization and minimize the computation complexity, we calculate the Fermat point of dense areas as the flight direction of mobile UAVs. In addition, we omit the slot subscript from (\ref{bp2}) for convenience.
\subsection{Beam Optimization}
Assuming that the load matrix and the traveling directions of the mobile UAVs are fixed, the beam optimization sub-problem can be expressed as
\begin{equation}
\begin{aligned}
&\mathop {{\rm{max}}}\limits_{\{ {{\bf{W}}_{u,k,n}}\} } \sum\limits_{u = 1}^U {\sum\limits_{k = 1}^K {{\rho _{u,k,n}}{{\bar R}_{u,k,\left. n \right|n - 1}}} } \label{bp}\\ 
&~~~~\;{\rm{s}}{\rm{.t}}{\rm{.}}~~(\ref{problem}{\mathrm{a}}),(\ref{problem}\mathrm{b}),(\ref{problem}\mathrm{c}),(\ref{problem}\mathrm{d})~\text{and}~(\ref{problem}\mathrm{h}).
\end{aligned}
\end{equation}
The interference of the objective function and the rank-one constraint in (\ref{problem}{c}) render the problem (\ref{bp}) non-convex. We relax the rank-one constraint, and (\ref{bp}) can be rewritten as
\begin{equation}
\begin{aligned}
&\mathop {{\rm{max}}}\limits_{\{ {{\bf{W}}_{u,k}}\} } \sum\limits_{u = 1}^U {\sum\limits_{k = 1}^K {{\rho _{u,k}}{\log _2}\!\!\left( {\!1 \!+\! \frac{{{\rm{tr}}({{\bar {\bf{h}}}_{u,k}}\bar {\bf{h}}_{u,k}^H{{\bf{W}}_{u,k}})}}{{\sum\limits_{i = 1,i \ne k}^K \!\!\!\!{{\rm{tr}}} ({{\bar {\bf{h}}}_{u,k}}\bar {\bf{h}}_{u,k}^H{{\bf{W}}_{u,i}}) \!+ \!\sigma _\mathrm{C}^2)}}} \!\right)} } \label{bp2}\\ 
&~~~~\;{\rm{s}}{\rm{.t}}{\rm{.}}~~(\ref{problem}\mathrm{a}),(\ref{problem}\mathrm{b}),(\ref{problem}\mathrm{d})~\text{and}~(\ref{problem}\mathrm{h}).
\end{aligned}
\end{equation}

The objective function of problem (\ref{bp2}) still remains non-concave, thus we apply SCA technology to approximate it as a concave function through iterative processes. Expand the  predicted achievable rate as follows
\begin{equation}
\begin{aligned}
\label{sca1}
\bar{R}_{u,k}&=\log_{2}{\left( {\textstyle \sum_{i=1}^K} \text{tr}(\bar{\mathbf{h}}_{u,k} \bar{\mathbf{h}}_{u,k}^H \mathbf{W}_{u,i})+\sigma_\mathrm{C}^2 \right)} \\
&~~~~-\log_{2}{\left( {\textstyle \sum_{i=1,i\neq k}^K} \text{tr}(\bar{\mathbf{h}}_{u,k} \bar{\mathbf{h}}_{u,k}^H \mathbf{W}_{u,i})+\sigma_\mathrm{C}^2 \right)}.
\end{aligned}
\end{equation}
For the $l$-th iteration, we implement first-order Taylor expansion to the second term of (\ref{sca1}) at $\{ \mathbf{W}_{u,k}^{(l)} \}$ and get
\begin{equation}
\begin{aligned}
\bar{R}_{u,k}\ge \log_{2}{\left( {\textstyle \sum_{i=1}^K} \text{tr}(\bar{\mathbf{h}}_{u,k} \bar{\mathbf{h}}_{u,k}^H \mathbf{W}_{u,i})+\sigma_\mathrm{C}^2 \right)} \\
\label{sca2}
~~~~-\log_{2}{\left( {\textstyle \sum_{i=1,i\neq k}^K} \text{tr}(\bar{\mathbf{h}}_{u,k} \bar{\mathbf{h}}_{u,k}^H \mathbf{W}_{u,i}^{(l)})+\sigma_\mathrm{C}^2 \right)}  \\
~~~~-{\textstyle \sum_{i=1,i\neq k}^K} \text{tr} \left(\mathbf{X}_{u,k}^{(l)}(\mathbf{W}_{u,i}-\mathbf{W}_{u,i}^{(l)})\right) \triangleq \tilde{R}_{u,k}^{(l)},
\end{aligned}
\end{equation}
where $\mathbf{X}_{u,k}^{(l)}$ is defined as
\begin{align}
\mathbf{X}_{u,k}^{(l)}=\frac{\log_{2}{(e)} \bar{\mathbf{h}}_{u,k}\bar{\mathbf{h}}_{u,k}^H}{{\textstyle \sum_{i=1,i\neq k}^K} \text{tr}(\bar{\mathbf{h}}_{u,k} \bar{\mathbf{h}}_{u,k}^H \mathbf{W}_{u,i}^{(l)})+\sigma_\mathrm{C}^2},
\end{align}
with $e$ as the natural constant.
Treat $\{\mathbf{W}_{u,i}\}$ in (\ref{sca2}) as $\{\mathbf{W}_{u,i}^{(l+1)}\}$, thus, problem (\ref{bp2}) evolves into the following SCA form
\begin{equation}
\begin{aligned}
&\mathop {{\rm{max}}}\limits_{\{ {{\bf{W}}_{u,k}}\} } \sum\limits_{u = 1}^U {\sum\limits_{k = 1}^K {{\rho _{u,k}}{{\tilde R}^{(l)}_{u,k}}} } \label{bp3}\\
&~~~~\;{\rm{s}}{\rm{.t}}{\rm{.}}~~(\ref{problem}\mathrm{a}),(\ref{problem}\mathrm{b}),(\ref{problem}\mathrm{d})~\text{and}~(\ref{problem}\mathrm{h}). 
\end{aligned}
\end{equation}
(\ref{bp3}) can be solved via convex optimization tools, such as CVX, a package for specifying and solving convex programs \cite{2008CVX}. 
Suppose that the optimal solution of the iteration is $\{ \mathbf{W}_{u,k}^{(l^\star)} \}$, the next step is to construct a rank-one solution by applying IRM. 

However, finding the rank-one solution of so many beamforming matrices at the same time results in a massive computational effort. Considering that each UAV can act as a computing terminal, we derive a distributed computing scheme, where each UAV individually works out its own beamforming matrix. We first convert the previous iteration of $\tilde{R}_{u,k}^{(l^\star)}$ as
\begin{equation}
\begin{aligned}
\begin{array}{l}\label{irm}
\tilde R_{u,k}^{({l^ \star } - 1)} = {\log _2}\left( {{\rm{tr}}({{\overline {\bf{h}} }_{u,k}}\overline {\bf{h}} _{u,k}^H{{\bf{W}}_{u,c}^{\dagger }})} \right.\\
~~~~~~\left. { + \mathop \sum \nolimits_{i = 1,i \ne c}^K {\rm{tr}}({{\overline {\bf{h}} }_{u,k}}\overline {\bf{h}} _{u,k}^H{{\bf{W}}_{u,i}^{({l^ \star } - 1)}}) + \sigma _\mathrm{C}^2} \right)\\
~~~~~~ - {\log _2}\left( {\mathop \sum \nolimits_{i = 1,i \ne k}^K {\rm{tr}}({{\overline {\bf{h}} }_{u,k}}\overline {\bf{h}} _{u,k}^H{\bf{W}}_{u,i}^{({l^ \star } - 1)}) + \sigma _\mathrm{C}^2} \right)\\
~~~~~~ - \mathop \sum \nolimits_{i = 1,i \ne k}^K {\rm{tr}}\left( {{\bf{X}}_{u,k}^{({l^ \star } - 1)}({{\bf{W}}_{u,i}^{({l^ \star })}} - {\bf{W}}_{u,i}^{({l^ \star } - 1)})} \right),
\end{array}
\end{aligned}
\end{equation}
where $c$ denotes a singled-out $c$-th user. For the third term in (\ref{irm}), if $c \ne k$, then when $i$ traverses to $c$, let ${{\bf{W}}_{u,c}^{\dagger }}$ replace ${{\bf{W}}_{u,i}^{({l^ \star })}}$. The IRM problems for each user are constructed as
\begin{align}
&~~~~~~~\mathop {{\rm{max}}}\limits_{ {{\bf{W}}_{u,c}^{\dagger }} } \sum\limits_{u = 1}^U {\sum\limits_{k = 1}^K {{\rho _{u,k}}{{\tilde R}^{({l^ \star } - 1)}_{u,k}}} } \label{irm1}\\
&~~~~~~~\;{\rm{s}}{\rm{.t}}{\rm{.}}~~{\rm{tr}}({{\bf{W}}_{u,c}^{\dagger }})={\rm{tr}}({{\bf{W}}_{u,c}^{( l^ \star)}}),\forall u,\tag{\ref{irm1}{a}}\\
&~~~~~~~~~~~~~~{{\bf{W}}_{u,c}^{\dagger }}\succeq0,{{\bf{W}}_{u,c}^{\dagger }} = ({{\bf{W}}_{u,c}^{\dagger }})^H,\forall u,\tag{\ref{irm1}{b}}\\
&~~~~~~~~~~~~~\;{{\tilde R}_{u,k}^{( l^ \star-1)}} \ge {\Gamma _{u,k}},\forall u,k,\tag{\ref{irm1}{c}}\\
&~~~~~~~~~~~~~~{\boldsymbol{a}^H}(o){{\bf{W}}_{u,c}}\boldsymbol{a}(o) + {B_k}{\rm{tr}}({{\bf{W}}_{u,c}}) \ge \nonumber\\
&~~~~~~~~~~~~~~\boldsymbol{a}^H({{\bar \theta }_{{u,c}}},{{\bar \varphi }_{{u,c}}}){{\bf{W}}_{u,c}}\boldsymbol{a}({{\bar \theta }_{{u,c}}},{{\bar \varphi }_{{u,c}}}),\nonumber \\
&~~~~~~~~~~~~~~~~~~~~~~~~~~~~~~~~~~~~~~~~~~\forall u,c,\forall o \in {\rm O}_{u,c}.\tag{\ref{irm1}{d}}
\end{align}
(\ref{irm1}) aims to solve the beamforming matrix ${{\bf{W}}_{u,c}^{\dagger }}$ of UAV $u$ designed for  user $c$, with the knowledge of other beamforming matrices $\{ \mathbf{W}_{u,k}^{(l^\star)} \left. \right| k \neq c\}$ and ${\rm{tr}}({{\bf{W}}_{u,c}^{( l^ \star)}})$. Accordingly, the computational problem is decomposed into $K$ sub-problems assigned to different UAVs. It can be proved that $\{{{\bf{W}}_{u,c}^{\dagger }}\}=\{ \mathbf{W}_{u,k}^{(l^\star)} \}$, and the last iteration of (\ref{bp3}) is equivalent to a set of (\ref{irm1})s corresponding to each user,respectively. Please see Appendix for details. 

In order to obtain the rank-one solution of (\ref{irm1}), let  $\{ \mathbf{M}_{u,c}^{(0)} \}$  represent $\{{{\bf{W}}_{u,c}^{\dagger }}\}$. We transform (\ref{irm1}) into another convex optimization problem, whose $q$-th iteration is as follows
\begin{align}
&~~~~\mathop {{\rm{max}}}\limits_{\{{\bf{M}}_{u,c}^{\left( q \right)}, {r^{\left( q \right)}}\}} \sum\limits_{u = 1}^\mathrm{U} {\sum\limits_{k = 1}^K {{\rho _{u,k}}\left( {\bar R_{u,k}^{(q)} + {w^q}{r^{\left( q \right)}}} \right)} } \label{irm2}\\
&~~~~~~{\rm{s}}.{\rm{t}}.{\rm{~~~tr}}({\bf{M}}_{u,c}^{\left( q \right)}) = {\rm{tr}}({\bf{W}}_{u,c}^{({l^ \star })}),\forall u,\tag{\ref{irm2}{a}}\\
&~~~~~~~~~~~~~{\rm{         }}{\bf{M}}_{u,c}^{\left( q \right)} \succeq 0,{\bf{M}}_{u,c}^{\left( q \right)} = {({\bf{M}}_{u,c}^{\left( q \right)})^H},\forall u,\tag{\ref{irm2}{b}}\\
&~~~~~~~~~~~~~{\rm{         }}\bar R_{u,k}^{(q)} \ge {\Gamma _{u,k}},\forall u,k,\tag{\ref{irm2}{c}}\\
&~~~~~~~~~~~~~{\rm{         }}{{\boldsymbol{a}}^H}(o){\bf{M}}_{u,c}^{\left( q \right)}{\boldsymbol{a}}(o) + {B_k}{\rm{tr}}({\bf{M}}_{u,c}^{\left( q \right)}) \ge \nonumber \\
&~~{\rm{         }}{{\boldsymbol{a}}^H}({{\bar \theta }_{u,c}},{{\bar \varphi }_{u,c}}){\bf{M}}_{u,c}^{\left( q \right)}{\boldsymbol{a}}({{\bar \theta }_{u,c}},{{\bar \varphi }_{u,c}}),\forall u,c,\forall o \in {{\rm{O}}_{u,c}},\tag{\ref{irm2}{d}}\\
&~~~~~~~~{\rm{         }}{r^{\left( q \right)}}{{\bf{I}}_{{M_t} - 1}} - {\left( {{\bf{V}}_{u,c}^{(q - 1)}} \right)^T}{\bf{M}}_{u,c}^{(q)}{\bf{V}}_{u,c}^{(q - 1)} \succeq 0,\tag{\ref{irm2}{e}}\label{irm2g}
\end{align}
where $\bar R_{u,k}^{(q)}$ is denoted by
\begin{equation}
\begin{aligned}
\begin{array}{l}\label{irm3}
\bar R_{u,k}^{(q)} = {\log _2}\left( {{\rm{tr}}({{\overline {\bf{h}} }_{u,k}}\overline {\bf{h}} _{u,k}^H{{\bf{M}}_{u,c}^{\left( q \right)}})} \right.\\
~~~~~~\left. { + \mathop \sum \nolimits_{i = 1,i \ne c}^K {\rm{tr}}({{\overline {\bf{h}} }_{u,k}}\overline {\bf{h}} _{u,k}^H{{\bf{W}}_{u,i}^{({l^ \star } - 1)}}) + \sigma _\mathrm{C}^2} \right)\\
~~~~~~ - {\log _2}\left( {\mathop \sum \nolimits_{i = 1,i \ne k}^K {\rm{tr}}({{\overline {\bf{h}} }_{u,k}}\overline {\bf{h}} _{u,k}^H{\bf{W}}_{u,i}^{({l^ \star } - 1)}) + \sigma _\mathrm{C}^2} \right)\\
~~~~~~ - \mathop \sum \nolimits_{i = 1,i \ne k}^K {\rm{tr}}\left( {{\bf{X}}_{u,k}^{({l^ \star } - 1)}({{\bf{W}}_{u,i}^{({l^ \star })}} - {\bf{W}}_{u,i}^{({l^ \star } - 1)})} \right).
\end{array}
\end{aligned}
\end{equation}
For the third term in (\ref{irm3}), if $c \ne k$, then when $i$ traverses to $c$, let ${\bf{M}}_{u,c}^{(q)}$ replace ${{\bf{W}}_{u,i}^{({l^ \star })}}$. It is worth noting that as the positive number $r^{(q)}$ approaches 0, (\ref{irm2g}) is the necessary and sufficient condition for ${\bf{M}}_{u,c}^{(q)}$ to be rank-one. ${\bf{V}}_{u,c}^{(q )}$ comprises of eigenvectors corresponding to the $M_t -1$ smaller eigenvalues of ${\bf{M}}_{u,c}^{(q)}$.  When $w^q$, the weight coefficient, is  sufficiently large with the iteration, $r^{(q)}$ becomes sufficiently small, then (\ref{irm2}) output an approximate solution of  rank-one.
\begin{table}[t]
\centering
\renewcommand{\arraystretch}{1.2}
% \small
\caption{Simulation Parameters}
\begin{tabular}{c|c}
 \hline %draw line
\textbf{parameter} & \textbf{value}  \\
\hline\hline
Max transmit power of UAV $P_{Tmax}$ (dBm) \cite{10061429} & 30   \\
\hline
Carrier frequency $f_c$ (GHz) \cite{10061429}& 30    \\
\hline
Length of slot $\Delta T$ (s)  & 0.01    \\
\hline
Communication path loss at reference distance $\alpha$ (dB) & -61.9   \\
\hline
Radar path loss at reference distance $\beta$ (dB) & -61.9  \\
\hline
AWGN $\sigma^2$ (dB)& -83   \\
\hline
Flight speed of mobile UAV $v_{n}^\mathrm{U}$ (m/s) & 30   \\
\hline
\end{tabular}
\label{parameter}
\vspace{-0.3cm}
\end{table}

\subsection{Load Optimization}
In order to enhance the total achievable rate, load optimization redistributes the load matrix $\textbf{P}_n$ in each slot based on the location and motion of all UAVs. However, adopting a new load matrix means that the beams have to be redesigned. Thereupon we propose a coalition game \cite{9293257}.

At the beginning, all UAVs and users co-establish an initial load matrix $\textbf{P}_1$ in accordance with the initial preference. Each user selects the nearest UAV first. Then, the user who is farther away from the UAV that exceeds $K_\mathrm{max}$ selects another UAV or exits the network based on the distance and load from other UAVs. Then, the coalition $u$, which is defined as the set of users loaded by UAV $u$ together with UAV $u$, is denoted by $A_u$. The coalition game is given by $(\mathcal{A},R^{\text{total}},\mathcal{U},\mathcal{K})$, where $\mathcal{A} = \{ A_1,A_2,\dots, A_U \}$ is the coalition set and $R^{\text{total}}$ is the set of achievable rates. The achievable rate of UAV $u$ on coalition $A_u$ is represented as
\begin{align}
    R_u^{\text{total}}\left( {{A_u}} \right) = \sum\limits_{k \in {A_u}}\!\! {{{\bar R}_{u,k}}}.
\end{align}

When users are transferred from a high-load coalition to a low-load coalition, the total achievable rate of the network is potentially improved due to increased resource utilization of low-load UAVs. Based on the utilitarian order of \cite{5230848}, user $k$ in coalition $A_i$ is transferred to $A_j$, $(i,j\in \mathcal{U})$  if the following condition is satisfied
\begin{align}\label{tranferRule}
    R_i^{\text{total}}\left( {{{A}_i'}} \right) + R_j^{\text{total}}\left( {{{A}_j'}} \right) > R_i^{\text{total}}\left( {{A_i}} \right) + R_j^{\text{total}}\left( {{A_j}} \right)
\end{align}
 where ${{{A}_i'}}$ and ${{{A}_j'}}$ represent the coalition after the transfer. According to \cite{2009A}, the coalition game  eventually reaches a stable matching. Note that since we impose restrictions on the maximum communication radius of each UAV, the possible transferring user do not have to traverse all coalitions. In addition, the water-filling method \cite{9171304} is employed to simplify the beam design. %The coalition game is summarized in Algorithem 2.

% \begin{algorithm}[!ht]
%     \renewcommand{\algorithmicrequire}{\textbf{Input:}}
% 	\renewcommand{\algorithmicensure}{\textbf{Output:}}
% 	\caption{Coalition game}
%     \label{power}
%     \begin{algorithmic}[1] % 控制是否有序号
%         \REQUIRE  Initial preference and initial load matrix $\textbf{P}_1$, convergence threshold $\epsilon_3$. % input 的内容
%         \STATE Calculate achievable rate $R_u^{\text{total}}\left( {{A_u}} \right)$ of each UAV  .
%         \REPEAT
%             \STATE Transfer user $k$ that satisfy (\ref{tranferRule}).
%             \STATE Update load matrix \textbf{P}.
%             \STATE Update the increment $\Delta R$ of total achievable rate.
%         \UNTIL $\Delta R<\epsilon_3$
% 	    \ENSURE Latest load matrix \textbf{P}; % output 的内容
%     \end{algorithmic}
% \end{algorithm}
\subsection{Mobile UAVs Direction Planning}
The direction planning of mobile UAVs faces similar difficulties as the load optimization. That is to say, after the location of a mobile UAV changes, it needs to face load optimization and beam design again to determine the total achievable rate. Such direction planning is both complex and prone to short-sightedness,  as it often results in local optimal. This undermines the objective of alleviating load pressure. Since the key factor affecting the total achievable rate is the location of all UAVs and users, we turn the problem into finding the optimal global location for the mobile UAVs, taking this location as the direction of flight. We consider a Fermat point search algorithm based on the fminsearch function of MATLAB, which further simplifies the search for the optimal location. In detail, mobile UAVs prioritize users that are excluded from the network because of exceeding the load limit of their nearest UAVs. If all detected users are integrated into the network, the mobile UAV then seeks Fermat point within its coalition.

% The direction planning of mobile UAVs faces similar difficulties as the load optimization. That is to say, after the location of a mobile UAV changes, it needs to face load optimization and beam design again to determine the total achievable rate. Such direction planning is both complex and prone to short-sightedness,  as it often results in local optimal. This undermines the objective of alleviating load pressure. Since the key factor affecting the total achievable rate is the location of all UAVs and users, we turn the problem into finding the optimal global location for the mobile UAVs, taking this location as the direction of flight. The optimal location can be solved by genetic algorithm \cite{10600135} or particle swarm optimization algorithm \cite{10600135}. 
% Nevertheless, the problem scale and time complexity associated with the above algorithms are significant. In order to reduce the computational complexity, we consider a Fermat point search algorithm based on the fminsearch function of MATLAB, which further simplifies the search for the optimal location to find the  Fermat point. In detail, mobile UAVs prioritize users that are excluded from the network because of exceeding the load limit of their nearest UAVs. If all detected users are integrated into the network, the mobile UAV then seeks Fermat point within its coalition.

\begin{figure*}[!ht]
\centering
\subfloat[]{\includegraphics[width=2in]{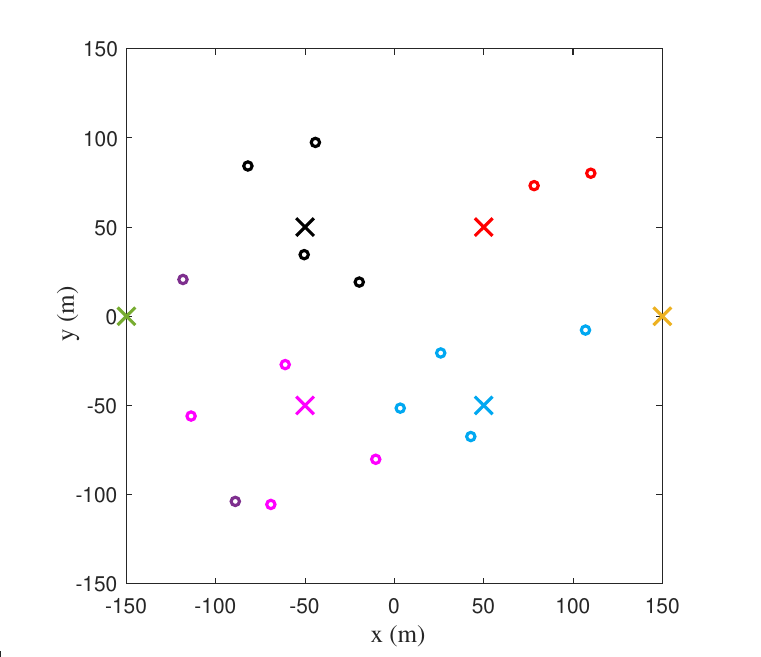}%
}
\hfil
\subfloat[]{\includegraphics[width=2in]{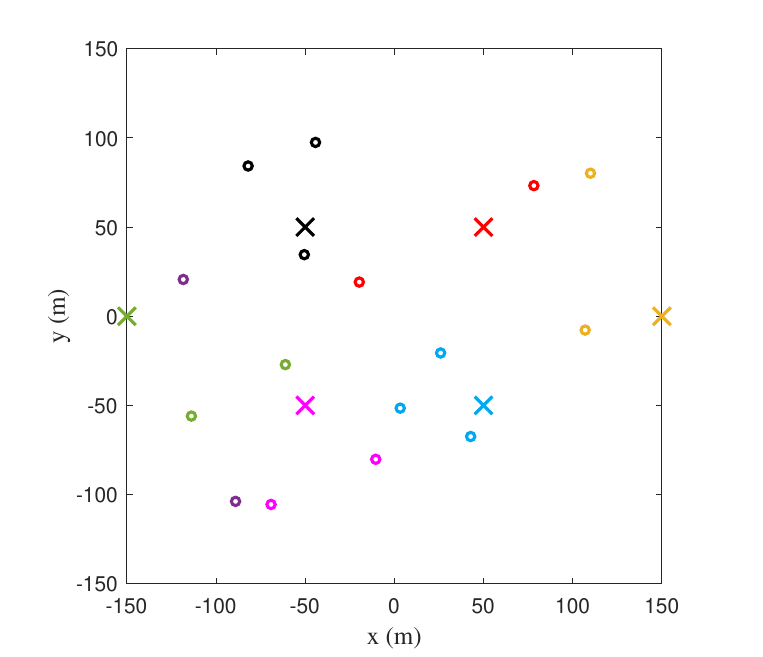}%
}
\hfil
\subfloat[]{\includegraphics[width=2in]{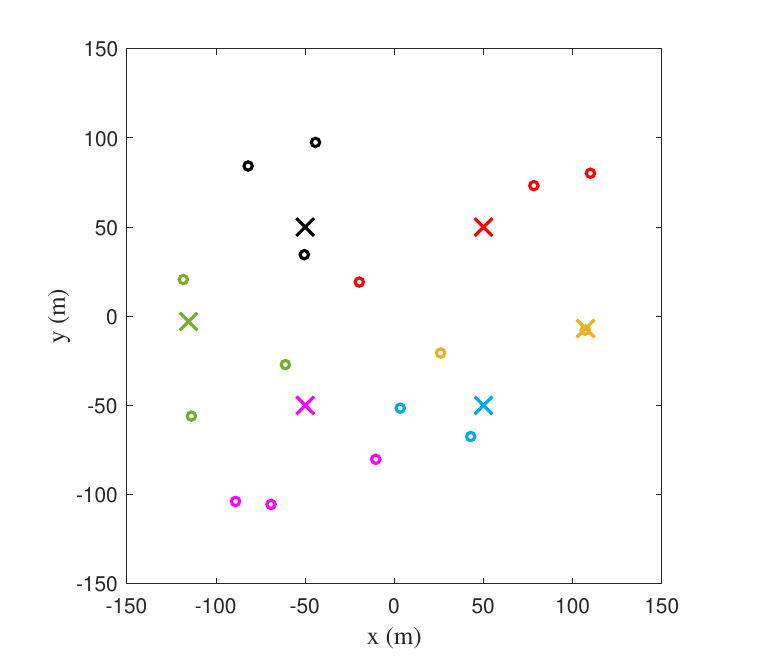}%
}
\hfil
\subfloat{\includegraphics[width=1in]{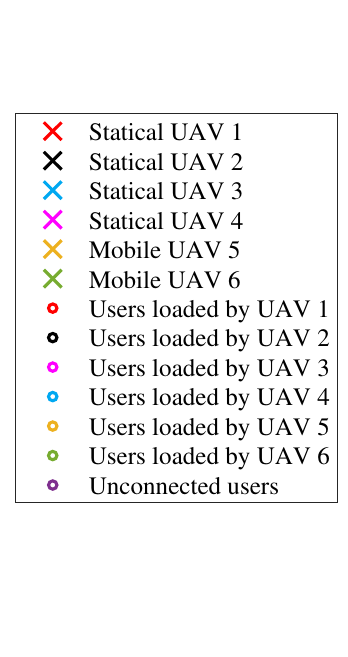}%
}
\caption{Bird's eye view of load change in multi-UAV network optimization process, where (a) denotes stage of initial load, (b) denotes stage of first load optimization, (c) denotes stage of load balance.}
\label{fig.3_loadchange}
\end{figure*}

\section{Numerical Results}
In this section, we evaluate the performance of the proposed method for both sensing and communication. Unless otherwise specified, the initial parameters are as follows: assuming maximum sensing/communication radius as $100$m, the locations of mobile users are distributed randomly and evenly in the ground area of $240$m $\times 240$m, the height of UAV is set to $h=100$m , coordinates of four static UAVs, numbered 1 to 4, are [$50$m, $50$m], [$50$m, $-50$m], [$-50$m, $50$m] and  [$-50$m, $-50$m], and coordinates of two mobile UAVs, numbered 5 and 6, are [$150$m, $0$m] and  [$-150$m, $0$m]. Moreover, we also set $\sigma^2 = \sigma^2_C$, and the noise power spectral density is considered as $-170$dBm/Hz with bandwidth $B_\mathrm{bw}=500$MHz. Both the transmit and receive arrays for the UAV are $M_t = M_r = 6\times6$, and the antenna spacing is configured as $\Delta d = \lambda/2$. According to \cite{9171304}, the variances for the evolution model should be small enough. It is set as follows: $\sigma_{\tilde \theta}=\sigma_{\tilde \varphi}=\sigma_{\tilde \varphi^v}=0.02^\circ$, $\sigma_{\tilde d}=0.02$m, $\sigma_{\tilde v}=0.025$m/s, $\sigma_{\tilde x^\mathrm{U}}=\sigma_{\tilde y^\mathrm{U}}=0.025$m and $\sigma_{\tilde z^\mathrm{U}}=0.01m$. In addition, we set the length of frame $N_L=10$ slots.
 Other simulation parameters are listed in Table \ref{parameter}.

\subsection{Performance of Load Optimization}
\begin{figure}[!h] %htbp
\centerline{\includegraphics[width=0.5\textwidth]{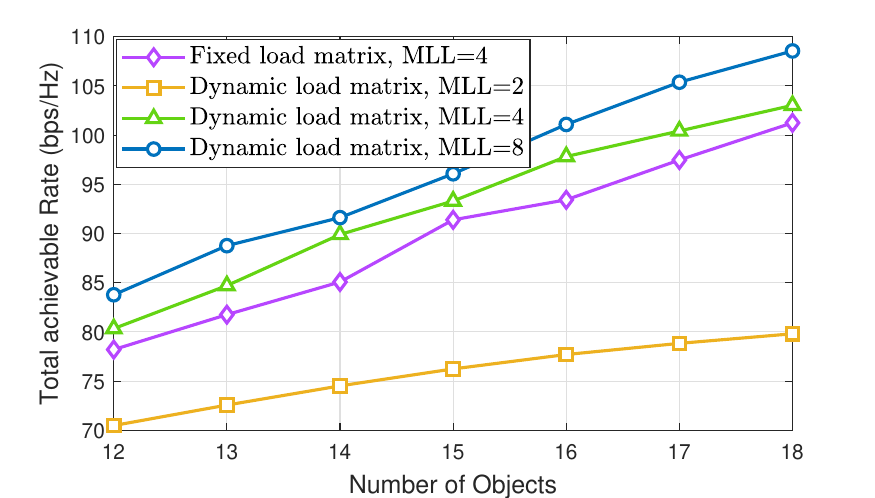}}
\caption{Total achievable rate under various numbers of users at slot $5$.}
\label{fig.1_userchange}
\vspace{-0.3cm}
\end{figure}
First, we discuss the performance of load optimization. Without direction planning,  Fig. \ref{fig.1_userchange} depicts the total achievable rate under various numbers of users at slot $5$. Generally, load balancing has been achieved in the fifth slot of the network. In the legend, the `Fixed load matrix' represents no load optimization, i.e., users connected with the nearest UAV, while the `Dynamic load matrix' indicates load optimization. It is evident that, irrespective of the scheme, as the number of users increases, both total achievable rates with and without optimization improve due to sufficient utilization of enhanced communication resources. In load optimization, as $K_{\text{max}}$ rises, communication performance also increase; however, the rate of increase diminishes as $K_{\text{max}}$ approaches the UAV network's maximum load capacity, leading to a reduced marginal effect. The total achievable rate for the `Fixed load matrix, $K_{\text{max}}=4$' is lower than that for the `Dynamic load matrix, $K_{\text{max}}=4$', yet significantly higher than that for the `Dynamic load matrix, $K_{\text{max}}=2$'. This indicates that load optimization enhances network performance, while $K_{\text{max}}$ can severely restrict it.

\begin{figure}[h] %htbp
\centerline{\includegraphics[width=0.5\textwidth]{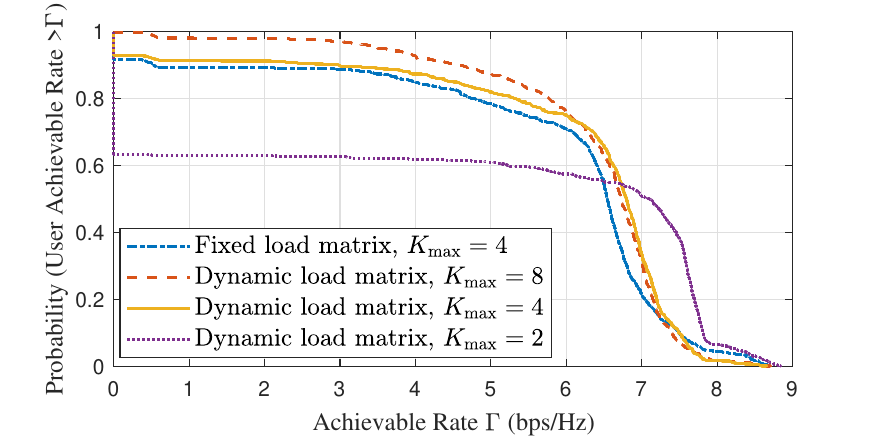}}
\caption{The CCDF curves of achievable rate at slot $5$.}
\label{fig.2_ccdf}
\vspace{-0.3 cm}
\end{figure}
To further elucidate our findings, we apply four different schemes to independently assess the achievable rate of each user in the fifth time slot. Following 100 Monte Carlo simulations, we derive the complementary cumulative distribution function (CCDF) for achievable rate, as depicted in Fig. \ref{fig.2_ccdf}. Under load optimization, the curves corresponding to different $K_{\text{max}}$  reveals that a lower $K_{\text{max}}$ correlates to a higher ratio of achievable rate approaching zero, suggesting that users are increasingly likely to be excluded from the network. At $K_{\text{max}}=2$, the number of users exhibiting medium achievable rate is minimized, with half of the users demonstrating achievable rate exceeding 7; this phenomenon arises from the reduced load pressure on each UAV, allowing for greater power allocation to the users. In contrast, at $K_{\text{max}}=4$ or $8$, the achievable rates are predominantly concentrated between 5 and 7, with disparities between the two schemes primarily reflecting the proportion of users with low achievable rate. Notably, when $K_{\text{max}}=8$, the network accommodates more users, which, due to power constraints, leads to an increased prevalence of low achievable rate. Overall, the total achievable rate for the load optimization scheme significantly surpasses that of the no load optimization scheme, indicating a higher proportion of users achieving moderate to high achievable rate. %Conversely, users with extreme high rate are more commonly found in the no load optimization scheme. %The above analysis underscores the effectiveness and equity of load optimization in enhancing network communication.%When $K_{\text{max}}=4$, we observe performance both with/without load optimization.%In other words, large $K_{\text{max}}$ is more likely to produce poor service quality.
\subsection{Performance of Mobile UAVs}

We then present the impact of mobile UAVs on networking. In Fig. \ref{fig.3_loadchange}, we analyze a scenario with 16 users and set the $K_{\text{max}}=4$, fixing the location of users for clarity. Specifically, $\times$ represents UAVs, while $\circ$  represents users. Users and UAVs of the same color indicate they belong to the same coalition, and purple $\circ$ denotes users that are not part of the network. The change depicted from left to right illustrates the initial load, first load optimization, and load balance. During the initial load, each user is assigned to the nearest UAV; once a UAV reaches its $K_{\text{max}}$, the corresponding user is excluded from the network. On the left side of Fig. \ref{fig.3_loadchange}, three coalitions have reached their load limit, resulting in two purple users being unable to connect to the network. Following the first load optimization, the integration of mobile UAV facilitates a more balanced distribution of load. After two frames, the strategic direction planning of the UAVs further alleviates the load and enhances the overall communication performance of the network. At this stage, all users are successfully integrated into the network, significantly improving both fairness and balance.

\begin{figure}[!h] %htbp
\centerline{\includegraphics[width=0.5\textwidth]{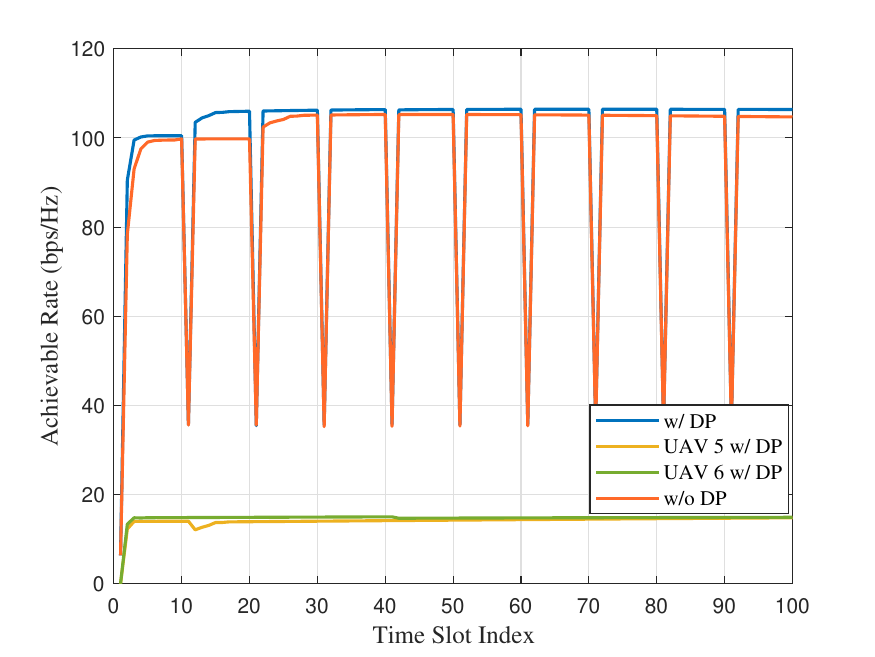}}
\caption{Comparison of network achievable rate with or without direction planning.}
\label{fig.4_directionopti}
\vspace{-0.3 cm}
\end{figure}
 Fig. \ref{fig.4_directionopti} illustrates the network achievable rate comparison of whether the mobile UAV adopts direction planning (i.e., DP in the figure). We consider a fixed number of 16 users, each randomly assigned a speed and direction of travel. For the network with DP, the total achievable rate reaches a stable value during the middle of the first frame, followed by a noticeable jump early in the second frame, where the  rate of mobile UAV 5 dips slightly. This dip occurs because, in the omnidirectional phase of the second frame, the load matrix is redistributed, allowing all users to join the network. Subsequently, the load matrix is optimized through coalition game, resulting in UAV 5 taking on a new load user. The newly added user also reaches stability after a brief period of beam optimization. For the network without DP, the rate of incorporating all users is delayed by one frame, as evidenced by the achievable rate variations observed in the initial three frames. Furthermore, while load balancing is attained, the total achievable rate of the scheme without directional planning remain inferior to those achieved with directional planning, with this disparity progressively increasing over time.
 %Additionally, it is worth noting that the achievable rate of UAV 6 decreases slightly in the fourth frame due to the changes in its load.

 %After an initial increase in the first two frames, the total achievable rate stabilizes, from which point onwards the load matrix remains unchanged. However, as the mobile UAVs continue to optimizes their directions, the total achievable rate experiences a slight increase. Specifically,
 
% \begin{figure}[!h] %htbp
% \centerline{\includegraphics[width=0.5\textwidth]{fig5_errorofdistance.pdf}}
% \caption{The influence of mobile UAVs on network sensing performance.}
% \label{fig.5_errorofdistance}
% \vspace{-0.3 cm}
% \end{figure}

% To highlight the significance of mobile UAVs for the network's sensing performance in terms of root mean squared error (RMSE), we create a scenario where two UAVs (one static and the other mobile) and five users, with $K_{\text{max}}$ set to 4. Initially, the static UAV cannot handle more than four users, resulting in the fifth user's distance sensing error being equivalent to that of omnidirectional detection, which is substantial. In the second frame, the mobile UAV assumes control of the fifth user, and as the directional beam is continuously optimized, the sensing information becomes increasingly accurate, quickly reducing the distance sensing error of user 5 to a lower range, as shown in Fig. \ref{fig.5_errorofdistance}.
\begin{figure}[!h]
\centering
\subfloat[]{\includegraphics[width=1.7in]{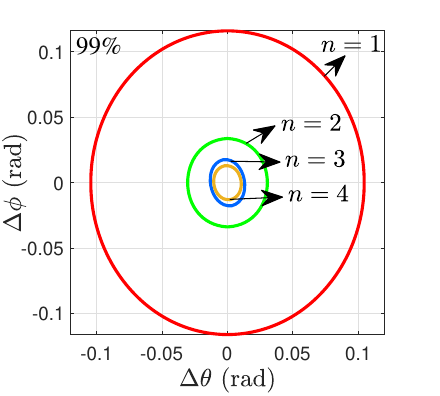}%
}
\hfil
\subfloat[]{\includegraphics[width=1.7in]{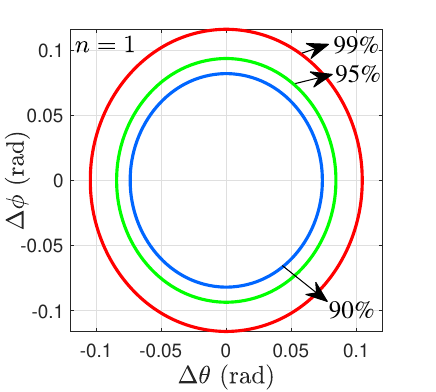}%
}
\caption{(a) Variation of confidence ellipse in different  slots at 99$\%$ confidence. (b) Variation of confidence ellipses for different confidence levels in the initial slot.}
\label{fig.6_confi}
\end{figure}

Furthermore, we evaluate the performance of beam optimization. As illustrated in Fig. \ref{fig.6_confi}, we set the static UAV at [0m, 0m] and the user  at [45m, 45$\sqrt{3}$m], with speed and travel direction being 30m/s and 45$^{\circ}$. With a confidence level of 99$\%$, the confidence ellipse, representing beam coverage, calculated for different slot $n$ is shown in Fig. \ref{fig.6_confi}(a), where the horizontal and vertical axes indicate errors in elevation and azimuth, respectively. The confidence ellipse for slot 1 is large due to poor omnidirectional sensing. In subsequent orientation phases, the ellipse becomes significantly smaller, stabilizing by $n=4$, when the elevation and azimuth errors are within 1 degree.

When focus  $n=1$, Fig. \ref{fig.6_confi}(b) depicts the confidence ellipse sizes at various confidence levels, demonstrating that larger ellipses correlate with higher confidence levels. Aligning the beam with the user more likely increases coverage, which, while enhancing robustness, results in higher power consumption and reduced gain.

\subsection{Performance of Beam Optimization}
\begin{figure}[!h]
\centering
\subfloat[]{\includegraphics[width=3.4in]{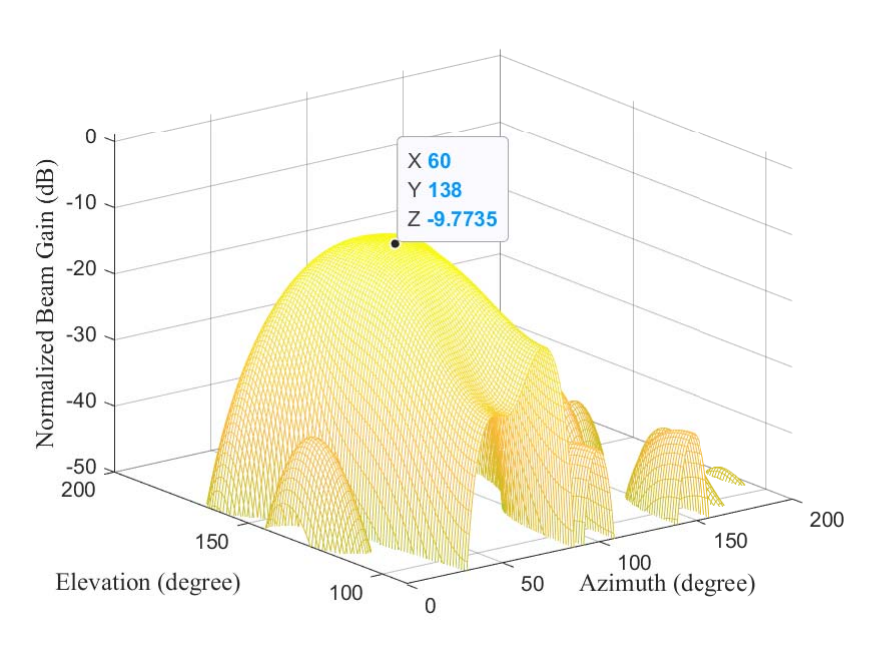}%
}
\hfil
\subfloat[]{\includegraphics[width=3.4in]{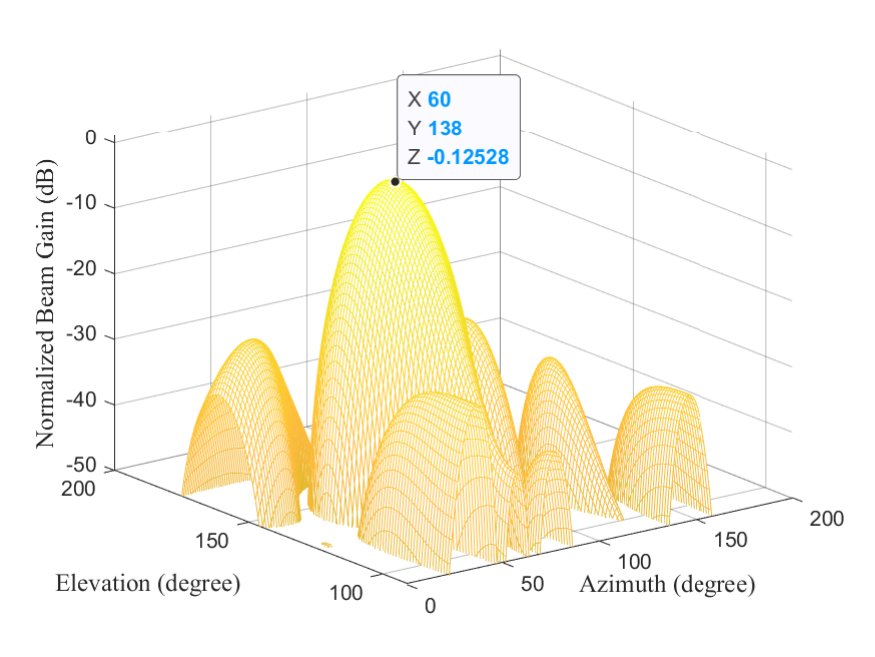}%
}
\caption{(a) UPA array form a beampattern in slot 1 on [138$^\circ$,60$^\circ$] direction user. (b) UPA array  form a beampattern in slot 3 on [138$^\circ$,60$^\circ$] direction user.}
\label{fig.9_beam}
\end{figure}

% \begin{figure}[h] %htbp
% \subfigure[]{
% \centerline{\includegraphics[width=0.5\textwidth]{fig9-1_beam1.pdf}}}
% \subfigure[]{
% \centerline{\includegraphics[width=0.5\textwidth]{fig9-2_beam3.pdf}}}
% \caption{(a) UPA array form a beampattern in slot 1 on [138$^\circ$,60$^\circ$] direction user. (b) UPA array  form a beampattern in slot 3 on [138$^\circ$,60$^\circ$] direction user.}
% \label{fig.9_beam}
% \vspace{-0.3cm}
% \end{figure}
Fig. \ref{fig.9_beam} illustrates the beam aggregation effect resulting from beam optimization. When a UAV is tasked with designing a beam for an user at elevation of 138$^\circ$, and azimuth of 60$^\circ$, the initial  slot shows a wider main lobe due to inaccurate sensing, leading to dispersed power and low beam gain. By the third  slot, the sensing information is sufficiently accurate, resulting in a significantly ``sharper'' main lobe, where power is focused in the target direction, greatly enhancing beam gain.

\section{Conclusion}
This paper presents a temporary sensing and communication scheme for multi-UAV networking. In the absence of prior information, we construct an omnidirectional time-division ISAC frame, allowing the transmitted ISAC signal to handle both communication and sensing tasks simultaneously. The sensing data extracted aids communication in the subsequent time slot. To alleviate communication load, we deploy statical UAVs to maintain the detection area and mobile UAVs to share load responsibilities, ensuring stable communication. We refined these tasks into optimization problems involving joint beam optimization, load optimization, and direction planning. Given the NP-hard nature of the problem, we divided them into three sub-problems and achieved  sub-optimal but high-quality solutions using improved SCA-IRM method, alliance games, and Fermat point search method. Numerical results demonstrate significant improvements in total achievable rate and sensing accuracy across beam design, load optimization, and direction planning compare to \cite{10806828}.

\appendix
\section*{The proof about the equivalence between last iteration of (\ref{bp3}) and set of (\ref{irm1}) corresponding to each user}
To avoid confusion, we prove that the last iteration problem (\ref{bp3}) with an objective function of $\tilde R_{u,k}^{({l^ \star } - 1)}$, denoted by (\ref{bp3}$\star$), is equivalent to the set of $K$ sub-problems of the form (\ref{irm1}), denoted by (\ref{irm1}$\star$).
\begin{theorem}\label{theorem1}
The proof of the equivalence of the optimization problems. $\iff$ The proof of the equivalence of the solutions of the optimization problems.
\end{theorem}
According to \cite{2004Convex}, we have Theorem \ref{theorem1}. In fact, observing the conversion of (\ref{bp3}$\star$) to (\ref{irm1}$\star$) above, it can learned that each sub-problem in (\ref{irm1}$\star$) is constructed by keeping the beamforming vector corresponding user $k$ as the variable, and the other optimal beamforming vectors as known quantities. Thus,
there is a Corollary \ref{corollary1}.
\begin{corollary}\label{corollary1}
Proving that the solutions of (\ref{bp3}$\star$) and (\ref{irm1}$\star$) are the same. $\iff$ proving that the optimal ${{\bf{W}}_{u,k}^{\dagger }}$ of the $k$-th sub-problem  is equal to the optimal $\mathbf{W}_{u,k}^{(l^\star)}$ of (\ref{bp3}$\star$), $\forall k$.
\end{corollary}
Noting that the optimization problems of (\ref{bp3}$\star$) and (\ref{irm1}$\star$) are both SDP convex problems, we can further deduce Corollary \ref{corollary2}.
\begin{corollary}\label{corollary2}
Proving that the solutions of (\ref{bp3}$\star$) and (\ref{irm1}$\star$) are the same. $\iff$ $\left(\ast \right):$ Prove that the optimal solution of the fixed partial variables in a convex programming problem does not change the optimal solution of the remaining variables.
\end{corollary}
To simplify the proof of $\left(\ast \right)$, the following multivariable convex programming problem is constructed
\begin{align}\label{sdp1}
\begin{array}{l}
\min {\rm{ }}f\left( \mathbf{X} \right)\\
{\rm{s}}{\rm{.t}}{\rm{. }}~{g_i}\left( \mathbf{X} \right) \le 0,\forall i,\\
~~~~~{h_j}\left( \mathbf{X} \right) = 0,\forall j,
\end{array}
\end{align}
where $\mathbf{X} = \left[ {{X_1},{X_2}} \right]$, and $f(\cdot)$, $g_i(\cdot)$, $h_j(\cdot)$ are all convex functions. Suppose $\mathbf{X}^\ast = \left[ {{X^\ast_1},{X_2^\ast}} \right]$, is the global optimal solution, then ${X_1^\ast}$ can be fixed and another optimization problem can be constructed as
\begin{align}\label{sdp2}
\begin{array}{l}
\min {\rm{ }}f\left( {X_1^ * ,{X_2}} \right)\\
{\rm{s}}{\rm{.t}}{\rm{. }}~{g_i}\left( {X_1^ * ,{X_2}} \right) \le 0,\forall i,\\
~~~~~{h_j}\left( {X_1^ * ,{X_2}} \right) = 0,\forall j.
\end{array}
\end{align}

\begin{theorem}\label{theorem2}
For convex programming, the Karush–Kuhn–Tucker (KKT) condition for a feasible solution is sufficient and necessary for the solution to be globally optimal.
\end{theorem}
According to \cite{1999Nonlinear}, we apply Theorem  \ref{theorem2} to construct the KKT condition for problem (\ref{sdp1}) as
\begin{align}
    \left\{\begin{aligned}
\nabla f\left({\mathbf{X}^\ast }\right)+\sum \lambda_{i} \nabla g_{i}\left(\mathbf{X}^\ast\right)+\sum \mu_{j} \nabla h_{j}\left(\mathbf{X}^\ast\right)  =0 \\
\mu_{j}  \geq 0 ;
\mu_{j} h_{j}\left(\mathbf{X}^\ast\right)  =0;
h_{j}\left(\mathbf{X}^\ast\right)  \leq 0
\end{aligned}\right.
\end{align}
% \begin{align}
%     \left\{\begin{aligned}
% \nabla f\left({\mathbf{X}^\ast }\right)+\sum \lambda_{i} \nabla g_{i}\left(\mathbf{X}^\ast\right)+\sum \mu_{j} \nabla h_{j}\left(\mathbf{X}^\ast\right) & =0 \\
% \mu_{j} & \geq 0 \\
% \mu_{j} h_{j}\left(\mathbf{X}^\ast\right) & =0 \\
% h_{j}\left(\mathbf{X}^\ast\right) & \leq 0
% \end{aligned}\right.
% \end{align}
Expanding $\mathbf{X}^\ast$, we have
\begin{align}\label{kkt}
    \left\{\begin{aligned}
\nabla f\left({X_1^\ast, X_2^\ast}\right)+\sum \lambda_{i} \nabla g_{i}\left(X_1^\ast, X_2^\ast\right)+~~~~~~~~\\
\sum \mu_{j} \nabla h_{j}\left(X_1^\ast, X_2^\ast\right)  =0 \\
\mu_{j}  \geq 0 ;
\mu_{j} \mathrm{h}_{j}\left(X_1^\ast, X_2^\ast\right)  =0 ;
h_{j}\left(X_1^\ast, X_2^\ast\right)  \leq 0
\end{aligned}\right.
\end{align}
% \begin{align}\label{kkt}
%     \left\{\begin{aligned}
% \nabla f\left({X_1^\ast, X_2^\ast}\right)+\sum \lambda_{i} \nabla g_{i}\left(X_1^\ast, X_2^\ast\right)+~~~~~~~~&\\
% \sum \mu_{j} \nabla h_{j}\left(X_1^\ast, X_2^\ast\right) & =0 \\
% \mu_{j} & \geq 0 \\
% \mu_{j} \mathrm{h}_{j}\left(X_1^\ast, X_2^\ast\right) & =0 \\
% h_{j}\left(X_1^\ast, X_2^\ast\right) & \leq 0
% \end{aligned}\right.
% \end{align}
Suppose that $X_2 = X_2^\ast$  is a feasible solution of problem (\ref{sdp2}), and its KKT condition is the same as (\ref{kkt}), then $X_2^\ast$ is the globally optimal solution of (\ref{sdp2}). This thus completes the proof.
\bibliographystyle{ieeetr}
\bibliography{books}

\end{document}